\documentclass[%
 reprint,superscriptaddress,
nobibnotes,
 amsmath,amssymb,
 aps,
 prb,
longbibliography,
]{revtex4-1}

\usepackage{siunitx}
\usepackage{caption}
\usepackage{subcaption}
\usepackage{bm}%
\usepackage{xcolor} 
\usepackage{graphicx}
\usepackage{float}
\usepackage{epstopdf}
\usepackage{afterpage}
\usepackage{setspace}
\usepackage{booktabs}
\usepackage[colorlinks=true,linkcolor=black,citecolor=black]{hyperref}

\usepackage{mhchem}            
\usepackage[utf8]{inputenc}     
\usepackage[T1]{fontenc}        
\DeclareUnicodeCharacter{0308}{\"}  

\usepackage{ragged2e}

\usepackage{upgreek}
\definecolor{customcolor}{RGB}{200, 0, 200}
\expandafter\ifx\csname package@font\endcsname\relax\else
 \expandafter\expandafter
 \expandafter\usepackage
 \expandafter\expandafter
 \expandafter{\csname package@font\endcsname}%
\fi
\begin{document}

\setstretch{1.0}
\title{A Combined Theoretical and Experimental Study of Oxygen Vacancies in
Co$_3$O$_4$ for Liquid-Phase Oxidation Catalysis}

\author{Amir Omranpour}
\email{\textcolor{black}{amir.omranpour@rub.de}}
\affiliation{Lehrstuhl f\"ur Theoretische Chemie II, Ruhr-Universit\"at Bochum, 44780 Bochum, Germany}
\affiliation{Research Center Chemical Sciences and Sustainability, Research Alliance Ruhr, 44780 Bochum, Germany}

\author{Lea K\"ammerer}
\affiliation{Faculty of Physics and Center for Nanointegration Duisburg-Essen (CENIDE), University of Duisburg–Essen, 47057 Duisburg, Germany}

\author{Catalina Leiva--Leroy}
\affiliation{Laboratory of Industrial Chemistry, Ruhr-Universit\"at Bochum, 44780 Bochum, Germany}

\author{Anna Rabe}
\affiliation{Faculty of Physics and Center for Nanointegration Duisburg-Essen (CENIDE), University of Duisburg–Essen, 47057 Duisburg, Germany}

\author{Takuma Sato}
\affiliation{Max Planck Institute for Chemical Energy Conversion, 45470 M\"ulheim an der Ruhr, Germany}

\author{Soma Salamon}
\affiliation{Faculty of Physics and Center for Nanointegration Duisburg-Essen (CENIDE), University of Duisburg–Essen, 47057 Duisburg, Germany}

\author{Joachim Landers}
\affiliation{Faculty of Physics and Center for Nanointegration Duisburg-Essen (CENIDE), University of Duisburg–Essen, 47057 Duisburg, Germany}

\author{Benedikt Eggert}
\affiliation{Faculty of Physics and Center for Nanointegration Duisburg-Essen (CENIDE), University of Duisburg–Essen, 47057 Duisburg, Germany}

\author{Eugen Weschke}
\affiliation{Helmholtz-Zentrum Berlin für Materialien und Energie (HZB), 12489 Berlin, Germany}

\author{Jean Pascal Fandr\'e}
\affiliation{Heterogeneous Catalysis and Sustainable Energy, Max-Planck-Institut für Kohlenforschung, Kaiser-Wilhelm-Platz 1, 45470 Mülheim an der Ruhr, Germany}

\author{Ashwani Kumar}
\affiliation{Heterogeneous Catalysis and Sustainable Energy, Max-Planck-Institut für Kohlenforschung, Kaiser-Wilhelm-Platz 1, 45470 Mülheim an der Ruhr, Germany}

\author{Harun T\"uys\"uz}
\affiliation{Heterogeneous Catalysis and Sustainable Energy, Max-Planck-Institut für Kohlenforschung, Kaiser-Wilhelm-Platz 1, 45470 Mülheim an der Ruhr, Germany}
\affiliation{Catalysis and Energy Materials Group, IMDEA Materials Institute, Calle Eric Kandel 2, 28906, Getafe, Madrid, Spain}

\author{Martin Muhler}
\affiliation{Laboratory of Industrial Chemistry, Ruhr-Universit\"at Bochum, 44780 Bochum, Germany}

\author{Heiko Wende}
\affiliation{Faculty of Physics and Center for Nanointegration Duisburg-Essen (CENIDE), University of Duisburg–Essen, 47057 Duisburg, Germany}

\author{J\"org Behler}
\affiliation{Lehrstuhl f\"ur Theoretische Chemie II, Ruhr-Universit\"at Bochum, 44780 Bochum, Germany}
\affiliation{Research Center Chemical Sciences and Sustainability, Research Alliance Ruhr, 44780 Bochum, Germany}

\date{\today}

\begin{abstract}
In the present work, we investigate oxygen vacancies (V$_\mathrm{O}$) in Co$_3$O$_4$, both in the bulk phase and under liquid-phase ethylene glycol oxidation, by combining theoretical and experimental techniques. Density functional theory calculations for bulk Co$_3$O$_4$ show that introducing an oxygen vacancy reduces two adjacent Co$^{3+}$ ions to Co$^{2+}$ and narrows the band gap. The newly formed Co$^{2+}$ ions adopt high-spin configurations in distorted octahedral sites and remain stable in this state in \textit{ab initio} molecular dynamics simulations at $300$ K.
Computed O and Co K-edge X-ray absorption spectra for ideal and vacancy-containing Co$_3$O$_4$
show excellent agreement with the experimental data and serve as references to analyze the liquid-phase ethylene glycol oxidation. The comparison with experimental O K-edge spectra of fresh and post-reaction catalysts shows that fresh samples resemble the vacancy-containing reference, whereas post-reaction spectra shift toward the ideal reference. These results suggest that under liquid-phase ethylene glycol oxidation conditions, Co$_3$O$_4$ becomes more oxidized rather than reduced, by refilling preexisting oxygen vacancies.
This is further supported by the observation that higher O$_2$ pressures increase the conversion and that the catalyst remains stable and active over several cycles.
\end{abstract}

\maketitle

\section{Introduction}\label{sec:Introduction}
Cobalt oxide is a mixed-valence transition metal oxide that has gained considerable attention due to its distinctive chemical, physical, and electronic characteristics. Among its various forms, the cobalt spinel phase Co$_3$O$_4$ has been extensively investigated owing to its outstanding performance in oxidation catalysis, particularly in selective hydrocarbon oxidation~\cite{waidhas2020secondary,hill2017site,finocchio1997ftir}.
The coexistence of Co$^{2+}$ and Co$^{3+}$ ions in Co$_3$O$_4$ gives rise to versatile redox activity, high catalytic performance, tunable electronic structure, and intriguing magnetic behavior~\cite{P6389,P7139,P7140,P7141,P6435,P6405,P6444,P6387,P7142}. These properties make Co$_3$O$_4$ an attractive material for a broad range of technological and industrial applications, including oxidation catalysis of alcohols~\cite{P6391}, water oxidation~\cite{feizi2019cobalt,jiao2009nanostructured}, methane combustion~\cite{hu2008selective}, and CO oxidation~\cite{xie2009low}, as well as in lithium-ion batteries and gas sensors~\cite{li2005co3o4}.

\begin{figure*}[ht!]
\centering
\includegraphics[width=0.8\textwidth, trim=0 55 0 110, clip=true]{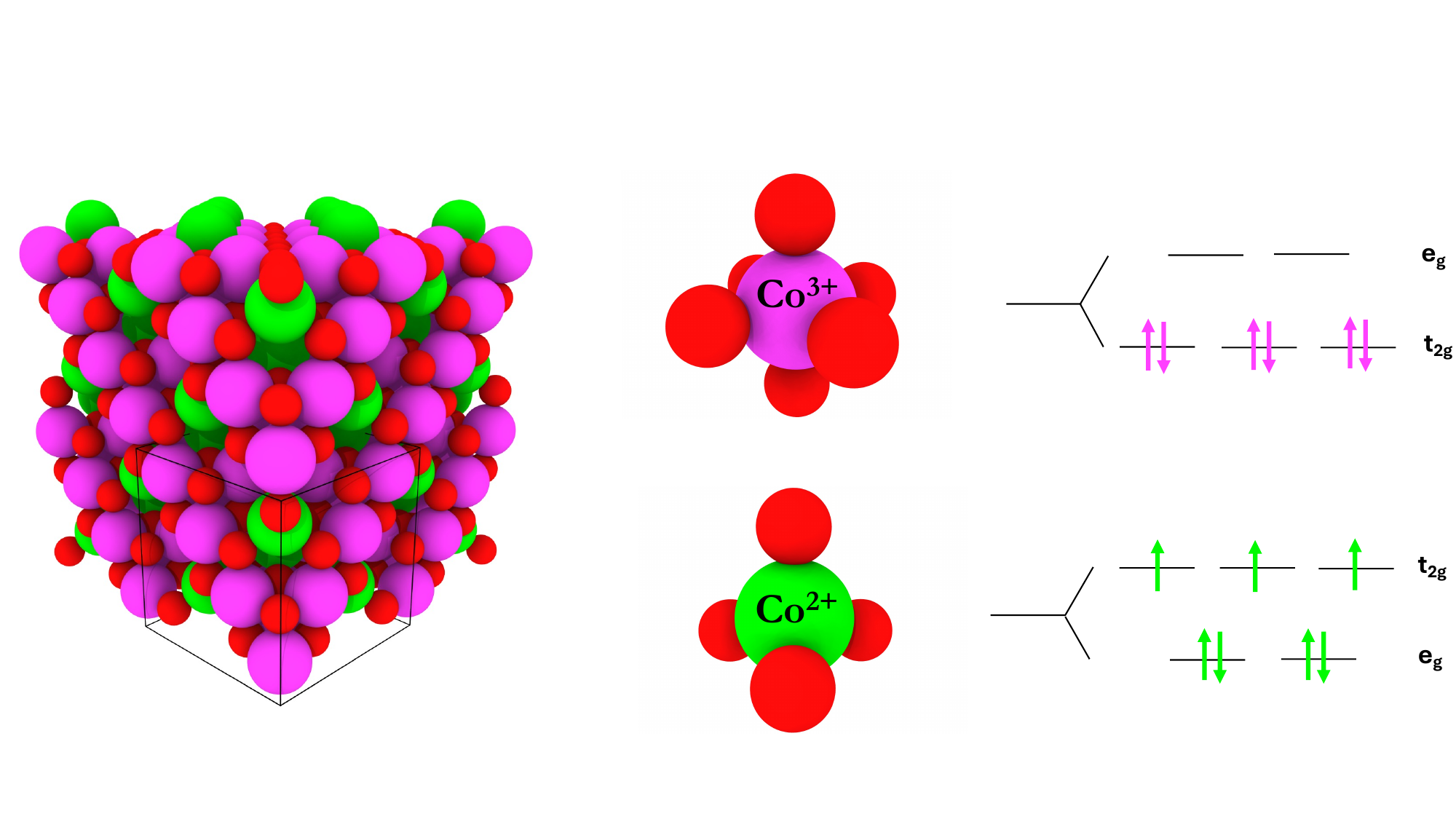}
\caption{\justifying
Co$_3$O$_4$ spinel ($2\times 2\times 2$) supercell. Co$^{2+}$ cations (green) occupy tetrahedral sites and Co$^{3+}$ cations (purple) occupy octahedral sites. The right side of the figure illustrates the corresponding crystal-field splittings: the lower-right diagram shows tetrahedrally coordinated Co$^{2+}$ in a $d^7$ configuration, and the upper-right diagram shows octahedrally coordinated Co$^{3+}$ in a $d^6$ configuration in the low-spin state.}
\label{fig:supercell}
\end{figure*}

Structurally, Co$_3$O$_4$ adopts a cubic $Fd\bar{3}m$ spinel structure in which cobalt exists in two oxidation states, Co\textsuperscript{2+} and Co\textsuperscript{3+}. The Co\textsuperscript{2+} ions occupy the tetrahedral interstitial sites, while Co\textsuperscript{3+} ions reside in the octahedral sites within the face-centered cubic (FCC) lattice formed by oxygen anions (Figure~\ref{fig:supercell}). The crystal field splits the five-fold degenerate $d$ orbitals of cobalt into two distinct energy levels. As a result, high-spin Co\textsuperscript{2+} exhibits three unpaired $d$ electrons, corresponding to an experimental magnetic moment of 3.26~$\mu_B$\cite{roth1964magnetic}, whereas Co\textsuperscript{3+} adopts a low-spin configuration with a nearly quenched magnetic moment (Figure~\ref{fig:supercell}). At ambient temperature, Co$_3$O$_4$ behaves as a paramagnetic semiconductor and becomes antiferromagnetic below 40~K\cite{roth1964magnetic}, mainly due to weak exchange interactions between neighboring Co\textsuperscript{2+} ions. Co$_3$O$_4$ is an intrinsic $p$-type semiconductor; its charge transport between $220$–$400~\mathrm{K}$ occurs predominantly via small-polaron hopping of holes, while in the $170$–$220~\mathrm{K}$ range it proceeds through variable-range hopping~\cite{cheng1998electrical,koumoto1981electrical}. The estimated experimental band gap of Co$_3$O$_4$ is approximately 1.6~eV\cite{kim2003optical,shinde2006supercapacitive}.

In heterogeneous catalysis, Co$_3$O$_4$ is recognized for its effectiveness in the oxygen evolution reaction and, more recently, in alcohol oxidation. Increasing attention has been directed toward performing such oxidation reactions in the liquid phase~\cite{najafishirtari2021perspective,P6391}, often in alkaline solution, which enables milder reaction conditions and improved selectivity \cite{leiva2025heat}.  Nevertheless, the precise influence of the aqueous environment on Co$_3$O$_4$ catalysis remains ambiguous, as both promoting and deactivating effects have been reported in computational~\cite{wang2012structural} and experimental~\cite{P6391,leiva2025heat} studies. Another open question is the actual state of Co$_3$O$_4$ under these conditions. Often, one starts from a “fresh’’ sample and assumes the catalyst remains relatively stable during the catalytic process. However, questions such as how the local structure, oxidation state, and magnetic state of Co$_3$O$_4$ change during the catalytic reaction have not yet been thoroughly answered. 

On the other hand, oxygen vacancies (V$_\mathrm{O}$) in Co$_3$O$_4$ are widely considered important for enhanced catalytic activity~\cite{liu2021surface,xu2016plasma,ferstl2015adsorption,fung2017general,montoya2011periodic,schellenburg2025mechanistic}, since they can increase the Co$^{2+}$/Co$^{3+}$ ratio and create sites that are more easily oxidized or reduced during the reaction cycle. However, most of these conclusions are based on gas-phase reactions or static models that do not address the stability of vacancy-induced Co$^{2+}$ species under liquid-phase oxidation conditions.

For these reasons, in this work, we combined theoretical and experimental approaches to (i) identify the local electronic and magnetic response to a single V$_\mathrm{O}$ in Co$_3$O$_4$, (ii) examine its stability at room temperature, and (iii) correlate these states with O K-edge measurements on fresh and post-reaction samples tested in the liquid-phase ethylene glycol oxidation, to further clarify the actual role of oxygen vacancies in this particular catalytic process. It should be noted that this work does not focus on the mechanistic aspects of ethylene glycol oxidation in the liquid phase, and the interested reader is referred to the literature for detailed mechanistic studies~\cite{nkou2025ethylene1,nkou2025ethylene2,leiva2025heat}.

The structure of this work is as follows. First, we present DFT results for ideal Co$_3$O$_4$ and for Co$_3$O$_4$ containing an oxygen vacancy, analyzing the changes in magnetic moments, density of states, and charge-density differences. We then use \textit{ab initio} molecular dynamics to test the finite-temperature stability of the vacancy-induced Co$^{2+}$ sites. Finally, we compare the calculated X-ray absorption spectra (O K-edge and Co K-edge) with the experimental spectra of fresh and post-reaction samples and discuss the implications for vacancy creation or healing under liquid-phase oxidation conditions.

\section{Computational Details}\label{sec:Computational}

All electronic structure calculations, \textit{ab initio} molecular dynamics simulations, and X-ray absorption spectroscopy calculations were carried out using the Vienna \textit{Ab initio} Simulation Package (VASP)~\cite{kresse1996efficient,kresse1996efficiency} (version~6.3.2) within the framework of spin-polarized density functional theory (DFT). Exchange--correlation effects and long-range dispersion interactions were described by the optPBE-vdW functional~\cite{perdew1996generalized,klimevs2009chemical,klimevs2011van}. On-site Coulomb interactions among the Co~$3d$ electrons were treated within the DFT+$U$ formalism of Dudarev~\textit{et~al.}~\cite{dudarev1998electron}, employing an effective Hubbard parameter of~$U_{\mathrm{eff}} = 2.43$~eV. We have shown previously that the aforementioned setting can reproduce the electronic and structural properties of Co$_3$O$_4$ in good agreement with  experimental and theoretical data~\cite{omranpour2024high}.

The interaction between valence electrons and ionic cores was represented by the Projector Augmented Wave (PAW) method~\cite{blochl1994projector}, as implemented by Kresse and Joubert~\cite{kresse1999ultrasoft,kresse1996efficiency}. Plane-wave expansions were performed with a kinetic energy cutoff of~500~eV. The Brillouin zone was sampled using a Monkhorst--Pack grid of~$5\times5\times5$~$\mathbf{k}$-points for bulk Co$_3$O$_4$. The bulk Co$_3$O$_4$ model contained 56~atoms in a cubic supercell of approximately~$8\times8\times8$~\AA$^3$ (see Ref.~\cite{omranpour2024high,omranpour2025insights}). Gaussian smearing of~0.1~eV was applied for partial occupancies, and non-spherical contributions inside the PAW spheres were included. Electronic self-consistency was reached when the total energy difference between successive iterations was below~$10^{-6}$~eV.

\textit{Ab initio} molecular dynamics (AIMD) simulations of the bulk Co$_3$O$_4$ were performed in the $NPT$ ensemble using a Langevin thermostat~\cite{parrinello1980crystal}. The target temperature was held at $T = 300\ \mathrm{K}$. A time step of $\Delta t = 0.5\ \mathrm{fs}$ was used for $N_{\mathrm{step}} = 100{,}000$ steps, yielding a total trajectory length of $50\ \mathrm{ps}$. Forces and stresses were evaluated at each step with the same electronic settings as above. The trajectories were generated at the $\Gamma$ point.

The O K-edge and Co K-edge X-ray absorption spectra were computed on top of the converged spin-polarized DFT+$U$ calculations described above. For both edges we employed a core-excited final-state approach within VASP~\cite{karsai2018effects}, in which the absorbing atom is replaced by a PAW potential containing a 1s core hole, while all other atoms retain the ground-state potentials. The same optPBE-vdW functional, $U_{\mathrm{eff}} = 2.43\,$eV, and plane-wave cutoff as in the ground-state calculations were used to ensure consistency. In order to minimize spurious interaction between periodic images of the core hole, the XAS was evaluated in the 56-atom Co$_3$O$_4$ supercell, and the core hole was placed on the site of interest, i.e., lattice O or Co in octahedral/tetrahedral coordination. 

Following recent theoretical XANES studies of Co$_3$O$_4$ at the O K-edge~\cite{P6387} and at the Co K-edge~\cite{douma2024electronic}, the calculated spectra were post-processed by convolution with an energy-dependent Lorentzian to account for core-hole lifetime broadening and inelastic losses, and an additional Gaussian of constant width to mimic the experimental energy resolution. A rigid energy shift was then applied so that the intense pre-edge/edge feature of the calculated spectra coincides with the corresponding experimental peak. The spectra were baseline-corrected and normalized to unity in a common post-edge energy window, consistent with the experimental normalization to the edge jump. For more details on these post-processing steps, the interested reader is referred to the above papers. For the Co K-edge, separate spectra were computed for tetrahedral Co$^{2+}$ and octahedral Co$^{3+}$ sites and combined in the stoichiometric ratio of 1:2 (Co$^{2+}_\mathrm{tet}$:Co$^{3+}_\mathrm{oct}$), using the same broadening, alignment, and normalization protocol for all ideal and vacancy-containing configurations.

\section{Experimental Details}\label{sec:Experiment}
\textbf{Liquid-phase ethylene glycol oxidation}
The experiments were performed in a batch reactor (Büchi) made of Hastelloy C-22, resistant to highly alkaline conditions. Catalytic reactions were conducted at 120°C under aerobic conditions (10 bar O$_2$) with 0.65 M KOH and 0.025 M or 0.325 M ethylene glycol (EG) for 6 h. Further details of Co$_3$O$_4$ synthesis procedure by co-precipitation, SBA-15 hard-templated and spray-flame can be found in earlier publications. \cite{dong2014co3o4,leiva2025heat,deng2017protocol,schulz2019gas} Standard characterization of the as-synthesized and post-catalysis of the hard-templated Co$_3$O$_4$ samples, and of further catalytic activity determination can be found in a previous study\cite{leiva2025heat}.

\textbf{X-ray absorption spectroscopy}
The O K-edge and Co $L_{2,3}$-edges X-ray absorption spectroscopy measurements were performed at the UE46-PGM1 \cite{Weschke2018} beamline at the synchrotron BESSY II. The measurements at the O K-edge have been carried out in the total electron yield (TEY) mode with linear horizontally polarized X-rays, while the Co $L_{2,3}$-edges (shown in Figure~S2) were measured with circular polarized X-rays. The absorption coefficient was calculated using the signal of the photocurrent divided by the monitor signal. The sample background was corrected by subtraction with a linear function applied in the pre-edge region of the data. Afterwards, the data was normalized with respect to the edge jump. The experimental data shown in Figure \ref{fig:O_XAS} are the result of two averaged scans for each sample.

The Co K-edge XAS measurements~\cite{fandre2025unveiling} were performed at the SAMBA beamline of the SOLEIL synchrotron, operating with an electron beam current of 450~mA. The incident energy was selected by a Si~(200) double crystal monochromator. Incident flux was ca. $1 \times 1010$~ph/s using a beam size of $1 \times 1$~mm. The Mn and Co K-edges were utilized to conduct measurements in transmission mode. The samples were prepared by diluting the corresponding powder with cellulose in a pellet ($\oslash = 13$~mm), followed by placement in the sample holder and sealing with Kapton tape. Calibration was performed using Mn and Co metal foils. The final spectra were normalized and processed using the Fastosh software (for more details, see Ref.~\cite{fandre2025unveiling}).

\textbf{Magnetometry}
The magnetic properties were recorded with the vibrating sample magnetometer (VSM) option of a Quantum Design PPMS DynaCool. Temperature-dependent M(T) magnetization curves were recorded with the zero field cooled - field cooled (ZFC-FC) protocol between 5 and 300\,K with an applied magnetic field of 0.1\,T.

\section{Results and Discussion}\label{sec:results}

\subsection{Oxygen Vacancies in Bulk Co$_3$O$_4$}\label{sec:General}

\subsubsection{DFT Calculations at 0 K}\label{sec:0K}

\begin{figure*}[ht!]
\centering
\includegraphics[width=1.0\textwidth, trim=0 0 0 110, clip=true]{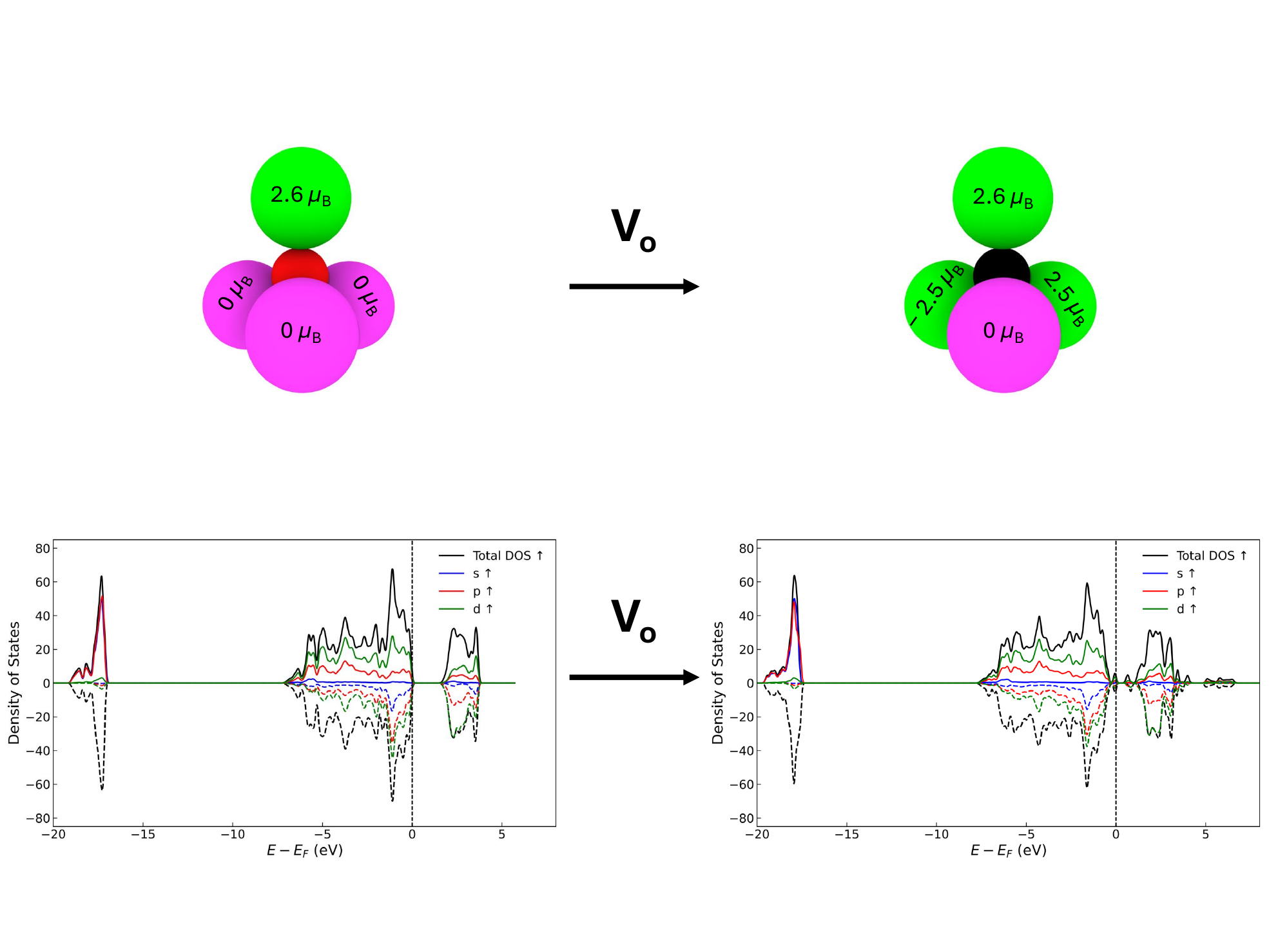}
\caption{\justifying
Local structural and electronic response of Co$_3$O$_4$ to the creation of an oxygen vacancy. Top: atomic configuration around the O site before (left) and after (right) removing the central lattice O atom. In the ideal case, the O atom is coordinated to three low-spin Co$^{3+}$ (purple, $0~\mu_\mathrm{B}$) and one high-spin Co$^{2+}$ (green, $2.6~\mu_\mathrm{B}$). After introducing the O vacancy, two of the neighboring Co$^{3+}$ ions are reduced to Co$^{2+}$, as indicated by their change to green and the appearance of finite magnetic moments, demonstrating local charge and spin redistribution. Bottom: spin-resolved, orbital-projected density of states (s, p, d) summed over all atoms. Solid lines denote the spin-up channel and dashed lines of the same color denote the spin-down channel (only spin-up is listed in the legend). In ideal Co$_3$O$_4$ a clear band gap is present. After the vacancy is created, additional Co 3$d$ character appears closer to the Fermi level, spin splitting becomes visible, and the band gap is narrowed.}
\label{fig:DOS}
\end{figure*}

\begin{figure*}[ht!]
\centering
\includegraphics[width=0.4\textwidth, trim=0 0 0 0, clip=true]{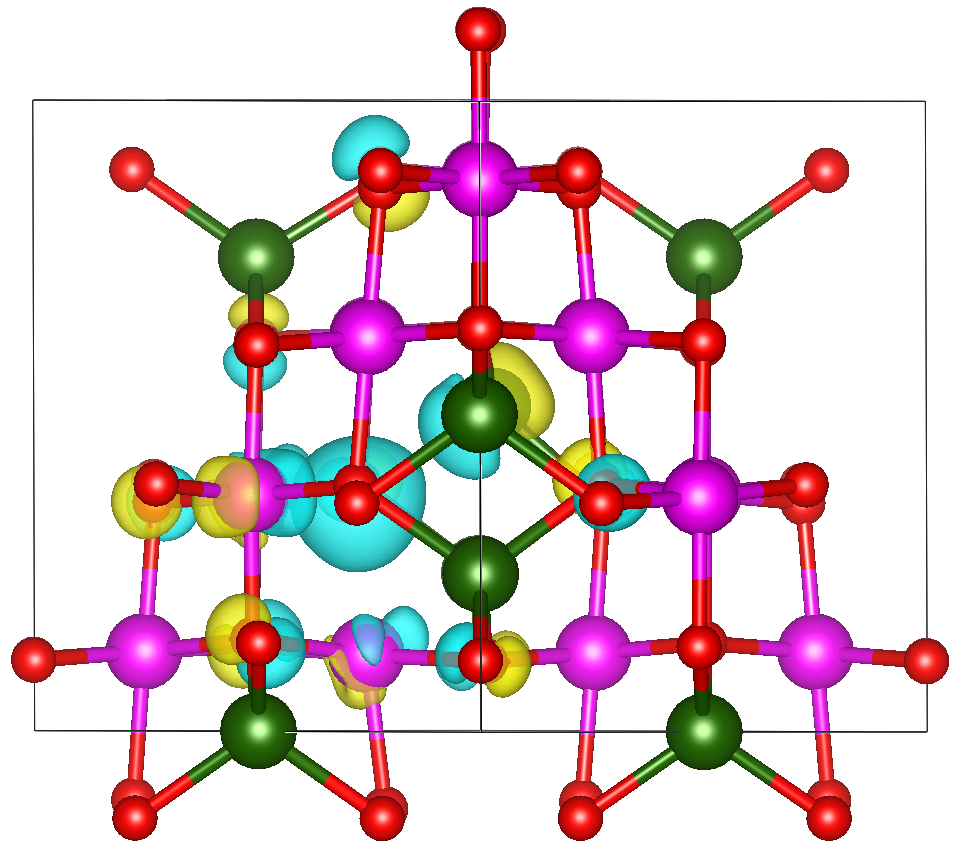}
\caption{\justifying
Charge density difference $\Delta \rho = \rho_{\mathrm{V_O}} - \rho_{\mathrm{ideal}}$ for Co$_3$O$_4$ with a single oxygen vacancy. Cyan isosurfaces indicate negative charge depletion at the vacancy site and along the former Co–O bonds, while yellow isosurfaces show negative charge accumulation on neighboring Co cations and nearby O atoms, showing redistribution of the electrons left behind by the removed oxygen.}
\label{fig:CDD}
\end{figure*}

First, DFT calculations for ideal Co$_3$O$_4$ were carried out and used as the reference throughout this work. They reproduce the expected antiferromagnetic ground state and gives a local magnetic moment of about $2.6~\mu_\mathrm{B}$ for the tetrahedral Co$^{2+}$ sites and $0~\mu_\mathrm{B}$ for the octahedral Co$^{3+}$ sites (see Table~S1 in the Supporting Information). The optimized lattice constant obtained with the present DFT setup is 8.156~\AA{} and the band gap is 1.61~eV, both in good agreement with experimental reports~\cite{zasada2015cobalt,liu1990high,shinde2006supercapacitive,roth1964magnetic} and consistent with previous DFT studies that are relevant to this work~\cite{douma2024electronic,P6387}. A more detailed validation of the accuracy of the current DFT settings is provided in Ref.~\citenum{omranpour2024high}.

Similar DFT calculations were performed for Co$_3$O$_4$ containing a single oxygen vacancy. A  comparison of the results with the ideal Co$_3$O$_4$ case is shown in Figure~\ref{fig:DOS}. In the upper panels, the local atomic environment around the oxygen site is shown before and after removing the O atom. In the ideal structure (left), one O atom is coordinated to four Co atoms: three are Co$^{3+}$ (purple) with $0~\mu_\mathrm{B}$ and one is Co$^{2+}$ (green) with $2.6~\mu_\mathrm{B}$. After introducing the oxygen vacancy (right), two of the Co$^{3+}$ cations adjacent to the vacancy are reduced to Co$^{2+}$, which are highlighted by the appearance of local magnetic moments (and are shown in green). This indicates that removing a single O atom redistributes charge and spin in the neighborhood of the vacancy, converting previously nonmagnetic, low-spin Co$^{3+}$ sites into high-spin Co$^{2+}$ sites. In addition, the two newly formed high-spin Co$^{2+}$ ions adopt opposite spin directions. Table~S2 in the supporting information provides the detailed orbital-resolved spin moments (s, p, d) and the resulting total local magnetic moment for every ion in the Co$_3$O$_4$ supercell. It may be worth adding that the existence of a finite X-ray magnetic circular dichroism (XMCD) signal, as shown in Figure~S2, may also be attributed to the uncompensated moments that arise when the Co$^{2+}$ ions adopt a high-spin configuration.

The lower part of Fig.~\ref{fig:DOS} shows the spin-resolved, orbital-projected density of states (DOS) for the two cases. The DOS is decomposed into $s$, $p$, and $d$ contributions. For each orbital the spin-up component is shown as a solid line, while the corresponding spin-down component is shown as a dashed line (only the spin-up curves are listed in the legend, but each has a dashed counterpart in the plot). For the ideal Co$_3$O$_4$ structure, the DOS displays a clear band gap and the $d$ states around the Fermi level are nearly spin-compensated, which is consistent with low-spin Co$^{3+}$. After introducing the oxygen vacancy, additional $d$-character appears closer to the Fermi level and the spin-up and spin-down $d$ curves no longer perfectly overlap; this is indicative of the reduction of neighboring Co$^{3+}$ to high-spin Co$^{2+}$ and the associated local spin polarization. At the same time, the onset of unoccupied states shifts so that the band gap is narrowed compared to the ideal case.

Figure~\ref{fig:CDD} shows the charge density difference (CDD) $\Delta \rho = \rho_{\mathrm{V_O}} - \rho_{\mathrm{ideal}}$ for Co$_3$O$_4$ after removing one lattice oxygen atom. In this color scheme, cyan denotes electron density \emph{depletion} and yellow denotes electron density \emph{accumulation}. A large cyan lobe is located at the position of the missing O atom and stretches along the former Co–O bonds, confirming that negative charge is removed from that region when the oxygen is taken out. The yellow isosurfaces appear mainly on the neighboring Co cations and along some of the remaining Co–O bonds, which shows where and how the electrons are redistributed. This indicates that due to the O vacancy, electron density is reallocated from the vacancy site itself and transferred to the adjacent Co–O units, thus locally reducing the nearby Co ions and polarizing the surrounding framework.

\subsubsection{The Role of Finite Temperature}\label{sec:AIMD}

\begin{figure*}[ht]
\centering
\begin{subfigure}{0.48\textwidth}
    \centering
    \caption{}
    \label{}
    \includegraphics[width=\textwidth, trim=0 0 0 0, clip=true]{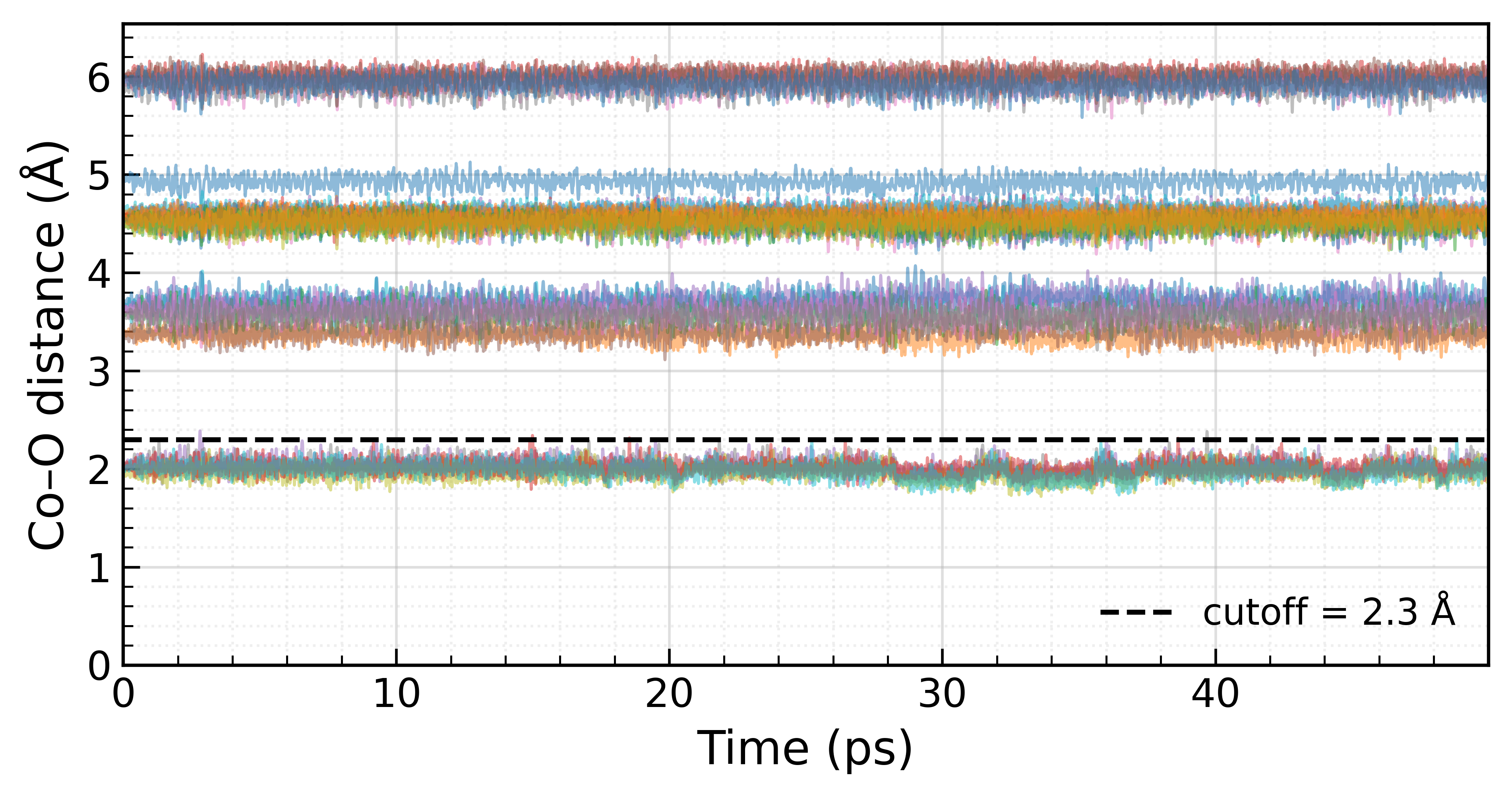}
\end{subfigure}%
\hfill
\begin{subfigure}{0.48\textwidth}
    \centering
    \caption{}
    \label{}
    \includegraphics[width=\textwidth, trim=0 0 0 0, clip=true]{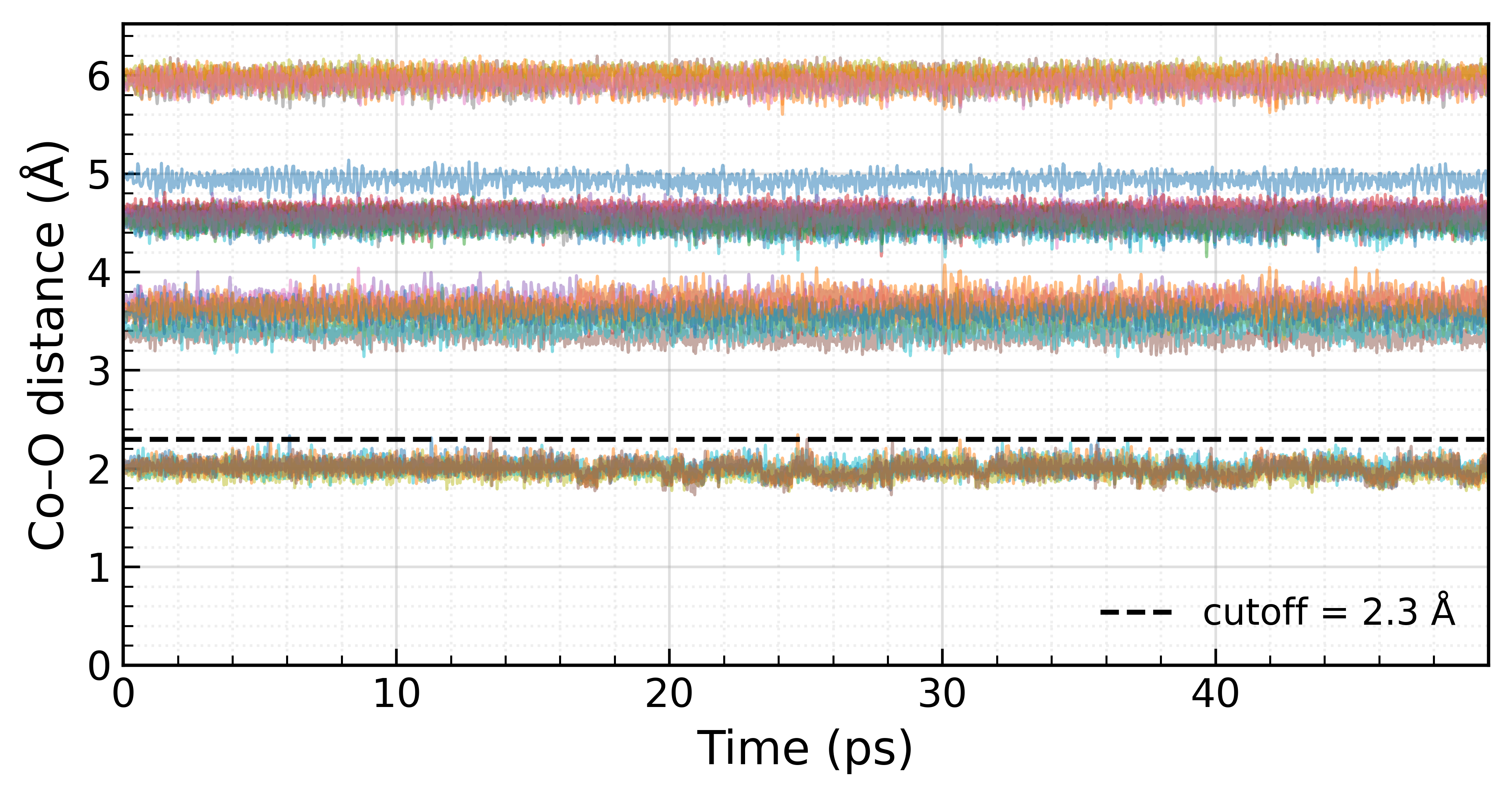}
\end{subfigure}

\vspace{0.4cm}

\begin{subfigure}{0.48\textwidth}
    \centering
    \caption{}
    \label{}
    \includegraphics[width=\textwidth, trim=0 0 0 0, clip=true]{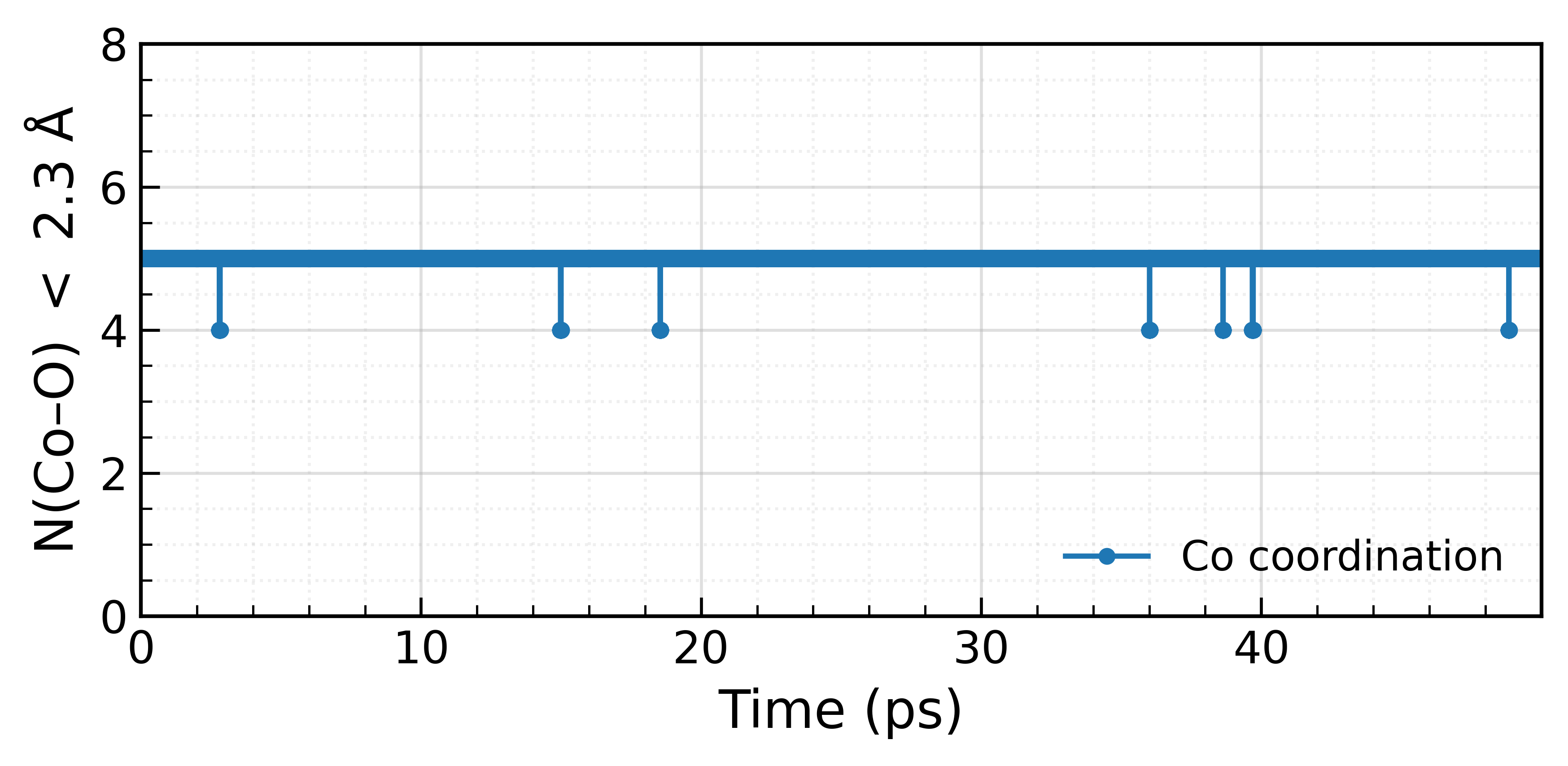}
\end{subfigure}%
\hfill
\begin{subfigure}{0.48\textwidth}
    \centering
    \caption{}
    \label{}
    \includegraphics[width=\textwidth, trim=0 0 0 0, clip=true]{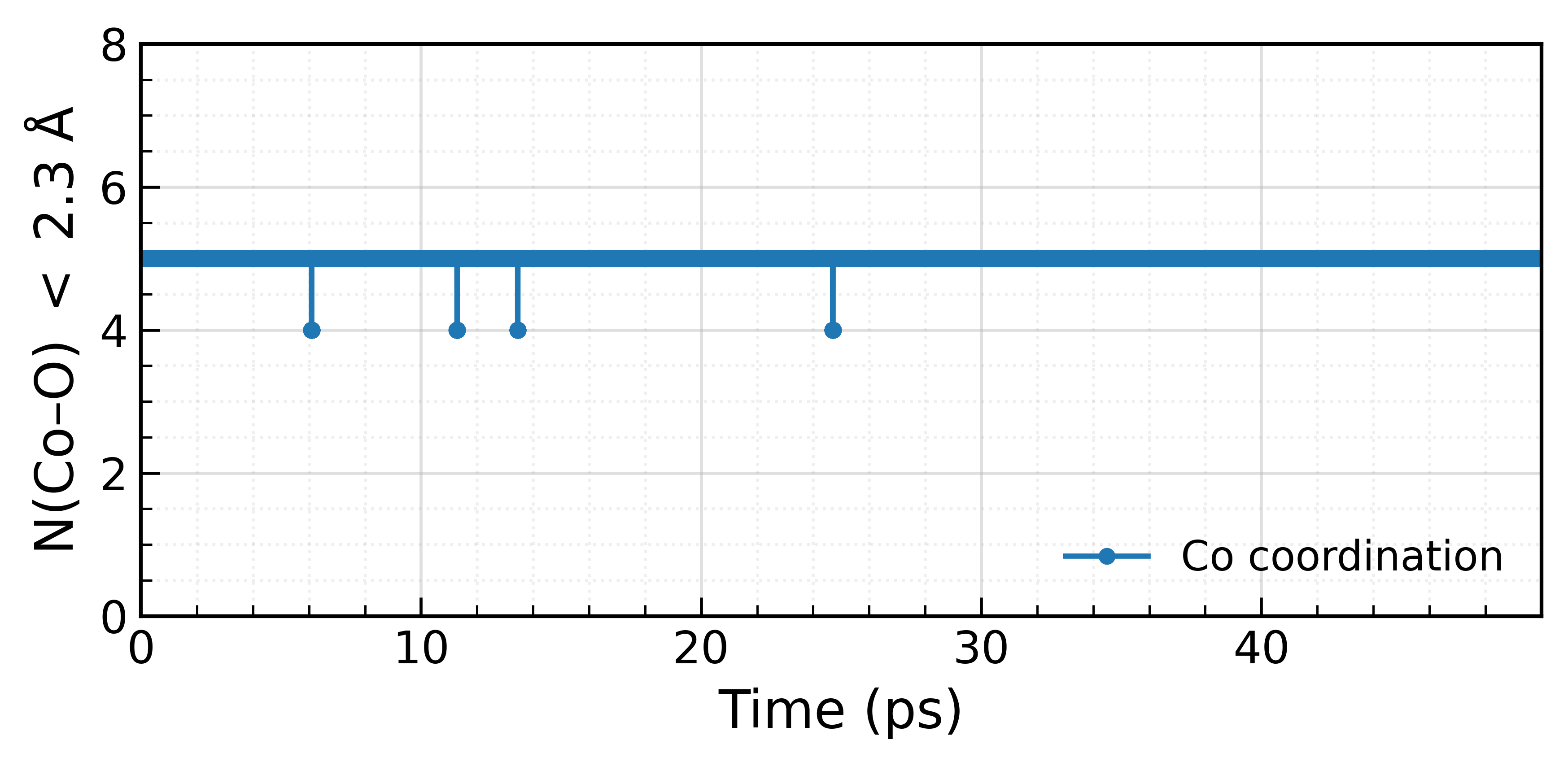}
\end{subfigure}

\vspace{0.4cm}

\begin{subfigure}{0.48\textwidth}
    \centering
    \caption{}
    \label{}
    \includegraphics[width=\textwidth, trim=0 0 0 0, clip=true]{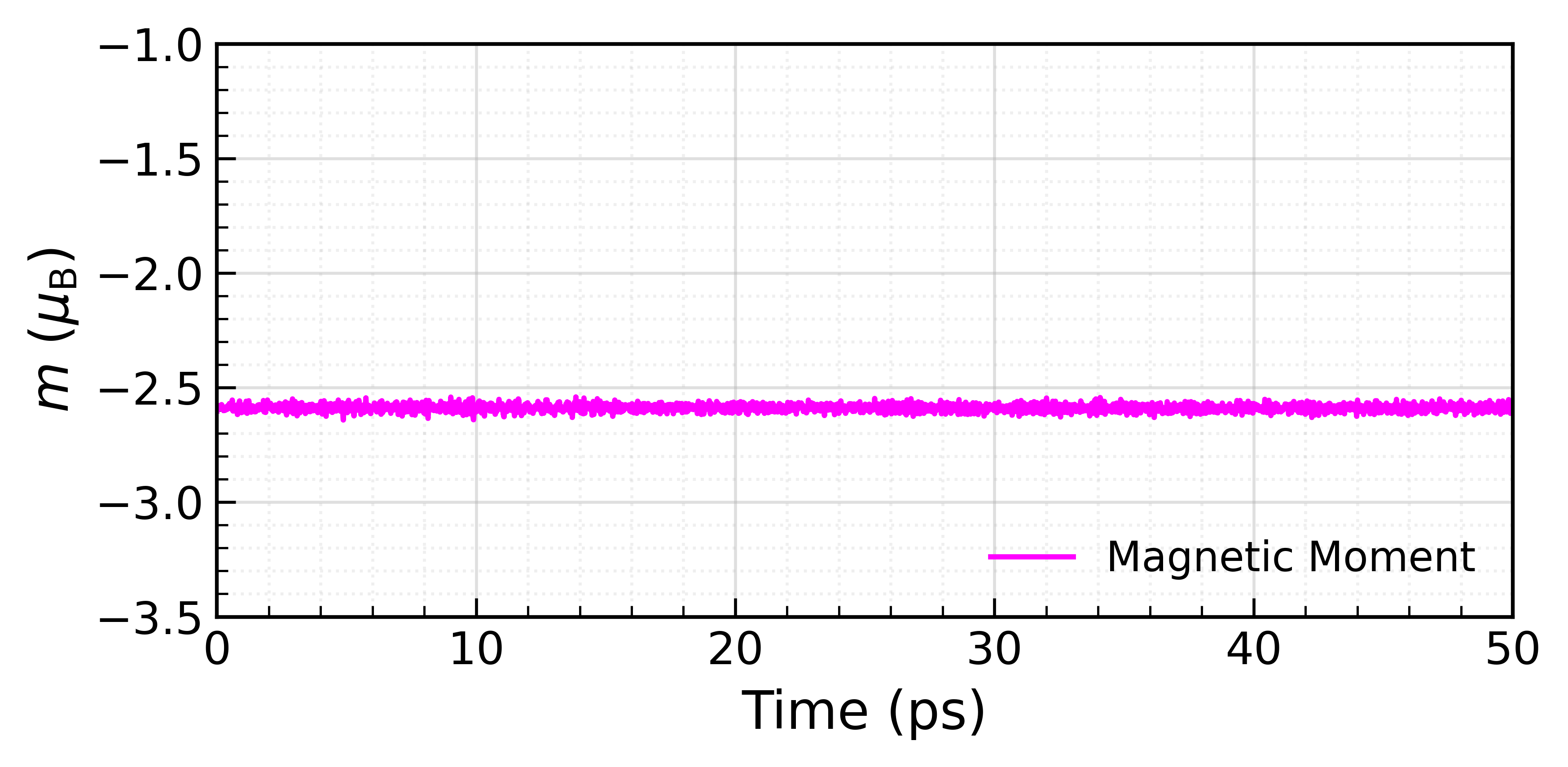}
\end{subfigure}%
\hfill
\begin{subfigure}{0.48\textwidth}
    \centering
    \caption{}
    \label{}
    \includegraphics[width=\textwidth, trim=0 0 0 0, clip=true]{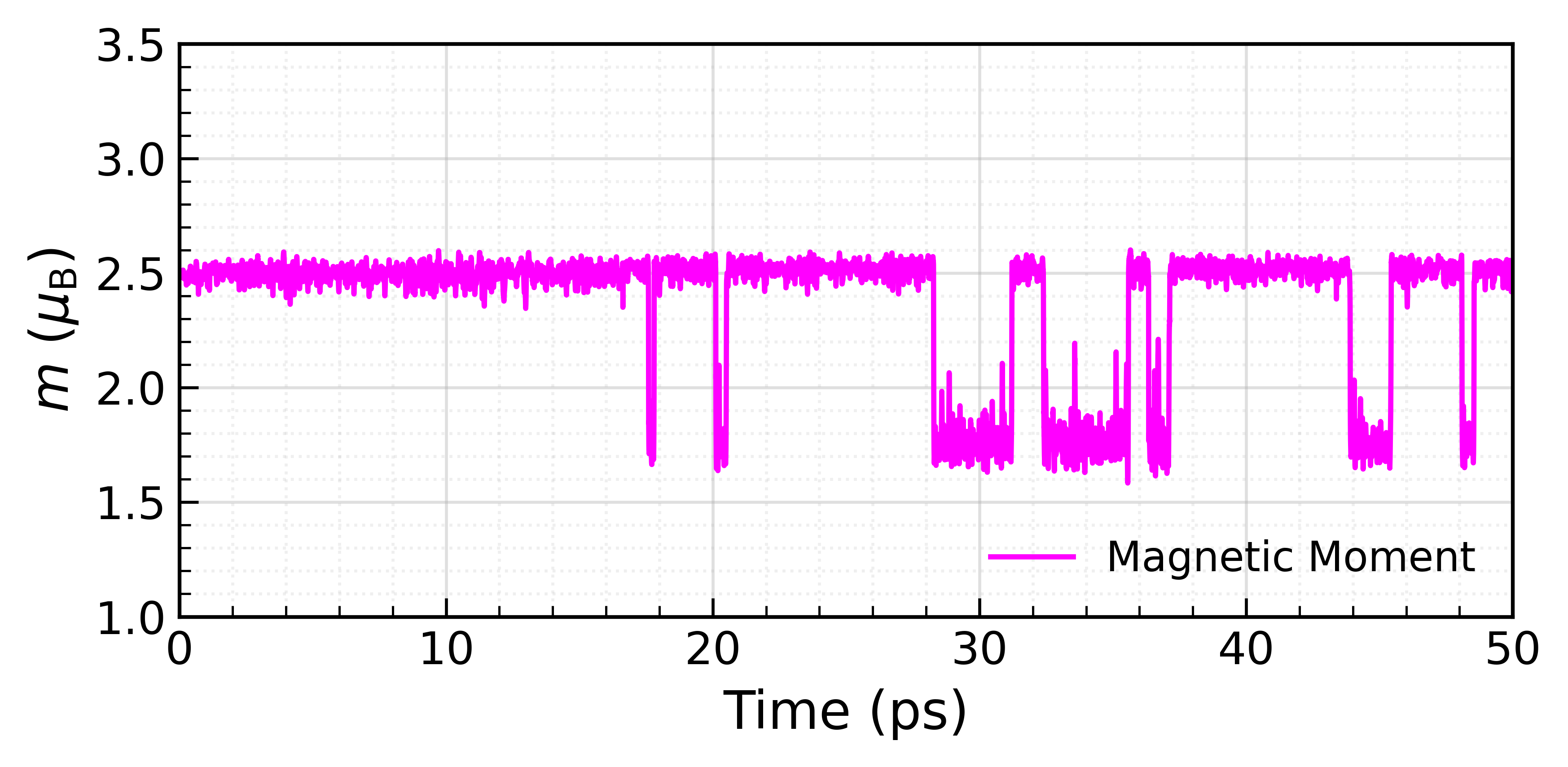}
\end{subfigure}

\caption{\justifying
Time evolution of the two Co ions next to the oxygen vacancy, followed over a $50~\mathrm{ps}$ \textit{ab initio} MD trajectory at $300~\mathrm{K}$. Panels~(a) and~(b) track all Co–O distances for Co~9 and Co~10. Each ion holds on to five short Co–O bonds in the range of $1.9$–$2.2~\text{\AA}$, while all remaining oxygen atoms are found beyond a distance of $3~\text{\AA}$. Neither site ever regains an octahedral environment; the five-fold coordination imposed by the vacancy remains intact. The coordination number plots in panels~(c) and~(d), obtained by counting O atoms within $2.3~\text{\AA}$ cutoff, show the same scenario: both Co$^{2+}$ centers fluctuate around a value of $5$, with only momentary drops to $4$ and never any increase to reach $6$. This configuration is therefore structurally stable at $300~\mathrm{K}$.
Panels~(e) and~(f) follow the magnetic moments. Co~9 stays essentially locked in its high-spin state at roughly $2.5~\mu_\mathrm{B}$ for the entire simulation. Co~10 shows occasional, very short-lived excursions down to $1.6$–$2.0~\mu_\mathrm{B}$, but always returns to the same high-spin value, and never displays anything resembling a spin collapse. Overall, these results show that the oxygen vacancy produces two Co$^{2+}$ ions that remain both structurally and magnetically robust over the full $50~\mathrm{ps}$ trajectory at $300~\mathrm{K}$.}
\label{fig:AIMD}
\end{figure*}

An important question for their role in catalysis is whether the Co$^{2+}$ species next to the oxygen vacancy in the optimized DFT structures remain stable at finite temperatures when the lattice is allowed to fluctuate. In the relaxed V$_\mathrm{O}$ structure the two Co atoms adjacent to the vacancy are five-fold coordinated (distorted octahedral) and carry a high-spin moment of about $2.5~\mu_\mathrm{B}$. To test the thermal stability of this configuration, we performed an \textit{ab initio} molecular dynamics simulation at 300~K starting from the relaxed V$_\mathrm{O}$ structure and propagated the system for 50~ps with a 0.5~fs time step. In the ideal Co$_3$O$_4$ unit cell there are eight Co$^{2+}$ sites, and here we label the two vacancy-induced Co$^{2+}$ as Co~9 and Co~10. The analysis is shown in Fig.~\ref{fig:AIMD}, where the left column (panels~(a), (c), (e)) corresponds to Co ion~9 and the right column (panels~(b), (d), (f)) corresponds to Co ion~10. 

Panels~(a) and~(b) report the time evolution of the distances between each of these Co atoms and \emph{all} O atoms in the unitcell. In both cases five Co--O distances stay clustered in the 1.9--2.2~\AA{} range for the entire 50~ps trajectory, while all other O atoms remain clearly farther away ($>3$~\AA). This indicates that neither of the two Co atoms regains a full sixfold octahedral environment, nor does it collapse to adopt a 
lower coordination. To quantify this, panels~(c) and~(d) show the corresponding coordination numbers as a function of time, obtained by counting all O atoms within 2.3~\AA{} of the Co ion. For both Co$^{2+}$ sites the coordination number is essentially constant at~5 throughout the trajectory, with only a few very short-lived drops to~4. These brief excursions coincide with a temporary elongation of one Co--O bond across the 2.3~\AA{} cutoff and do not persist. Importantly, we never observe an increase to 6, i.e., no spontaneous reoccupation of the vacant site or rearrangement into a regular octahedron occur at 300~K. We therefore conclude that the vacancy-induced, five-fold coordinated Co$^{2+}$ environment is structurally stable on the 50~ps time scale at room temperature, and that any further structural change would require overcoming a higher barrier than what thermal fluctuations at 300~K can provide.

Panels~(e) and~(f) track the local magnetic moments of the same two Co ions. Co~9 (panel~(e)) retains a nearly constant moment of $\sim -2.5~\mu_\mathrm{B}$ over the whole simulation, which is exactly the value found in the static DFT calculation and is characteristic of high-spin Co$^{2+}$. Co~10 (panel~(f)) behaves similarly for most of the simulation but exhibits several short intervals in which its moment drops to about $1.6$--$2.0~\mu_\mathrm{B}$. These episodes are transient and the moment always recovers to $\approx 2.5~\mu_\mathrm{B}$; at no point does the spin collapse to $\sim 0~\mu_\mathrm{B}$, i.e., we do not observe a conversion back to low-spin Co$^{3+}$. Because the number of Co--O neighbours does not change in a permanent way, these short drops in the magnetic moment cannot be attributed to a change of site or oxidation state. The more plausible explanation is that the Co ion remains Co$^{2+}$ in the same five-fold environment, but thermal motion at 300~K slightly perturbs the local bonds; consequently, the calculated moment briefly decreases before returning to its equilibrium value. Thus, these are small temperature-driven oscillations of the moment rather than a genuine Co$^{2+} \rightarrow$ Co$^{3+}$ transition or a structural rearrangement. Overall, the AIMD confirms that the oxygen-vacancy-induced Co$^{2+}$ sites are both \emph{structurally} and \emph{magnetically} robust at 300~K on the  50 picoseconds time scale. It may be worth noting that electron paramagnetic resonance (EPR) studies on Co$_3$O$_4$ platelets used for the partial oxidation of cinnamyl alcohol have likewise reported a surface-near distorted high-spin Co$^{2+}$ center, tentatively attributed to a distorted octahedral or tetrahedral site.\cite{schellenburg2025mechanistic}

\subsection{Liquid--Phase Ethylene Glycol Oxidation}\label{sec:EG}

\subsubsection{Catalytic Activity, Selectivity, Reusability, and the impact of \ce{O2} Pressure}\label{sec:Catalysis}

\begin{figure}[h]
\centering
\includegraphics[width=0.4\textwidth, trim=135 0 140 0, clip=true]{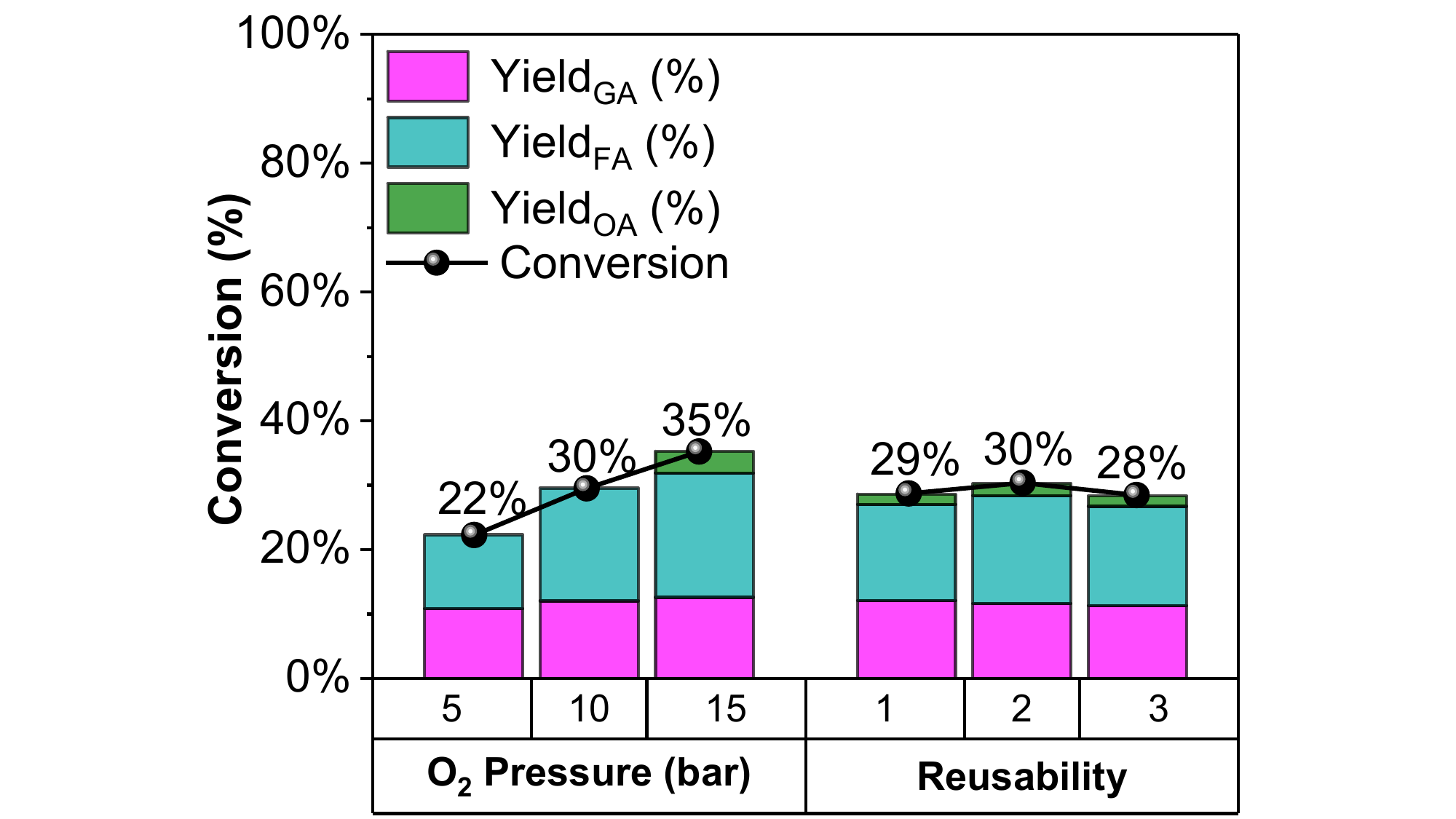}
\captionsetup{justification=justified}
\caption{\justifying
Effect of \ce{O2} pressure and catalyst reusability for the hard-templated Co$_3$O$_4$ sample in the liquid-phase oxidation of ethylene glycol. The left part shows the EG conversion and the corresponding yields/selectivities of glycolic acid (GA), formic acid (FA), and oxalic acid (OA) after $6~\mathrm{h}$ at $120\,^{\circ}\mathrm{C}$, $0.325~\mathrm{M}$ EG, and $0.65~\mathrm{M}$ KOH for \ce{O2} pressures of 5, 10, and 15~bar. The right part shows EG conversion and GA/FA/OA yields/selectivities for three consecutive runs at 10~bar \ce{O2} under the same conditions, which illustrates the stability and reusability of the hard-templated sample.}
\label{fig:O_pressure}
\end{figure}

Figure~\ref{fig:O_pressure} summarizes the relation between \ce{O2} pressure as well as catalyst reusability for the activity and selectivity of the hard-templated Co$_3$O$_4$ samples during liquid-phase ethylene glycol oxidation. The left part of the figure shows the ethylene glycol conversion and the yields of glycolic acid (GA), formic acid (FA), and oxalic acid (OA) after $6~\mathrm{h}$ at $120\,^{\circ}\mathrm{C}$, $0.325~\mathrm{M}$ EG, and $0.65~\mathrm{M}$ KOH for \ce{O2} pressures of 5, 10, and 15~bar. The total height of each bar corresponds to the overall EG conversion; the colored segments show the individual yields of GA, FA, and OA, respectively. By increasing the \ce{O2} pressure, the ethylene glycol conversion and the absolute yields of the oxidation products increase. The selectivity pattern changes only moderately: the glycolic acid (GA) yield does not increase substantially with increasing \ce{O2} pressure, while the relative contributions of formic acid (FA) and oxalic acid (OA) are enhanced, i.e., at higher \ce{O2} pressure, more strongly oxidized products are favored.

The right part of Figure~\ref{fig:O_pressure} shows the conversion and product yields for three consecutive reaction cycles, which are carried out at 10~bar \ce{O2} under the same alkaline conditions. Both overall conversion and the relative distribution among GA, FA, and OA remain mainly unchanged in consecutive cycles. Thus, it can be suggested that the hard-templated Co$_3$O$_4$ catalyst is not only active and selective but also structurally stable and reusable under the alkaline, \ce{O2}-rich liquid-phase reaction environment. Further details on the aforementioned experiments can be found in a recent publication\cite{leiva2025heat}.

\subsubsection{X-ray Absorption Spectroscopy}\label{sec:XAS}
\textbf{O K-edge X-ray Absorption Spectroscopy:}

\begin{figure*}[ht]
\centering
\begin{subfigure}{0.48\textwidth}
    \centering
    \caption{Experiment (Sample HT)}
    \label{fig:O_XAS_HT_exp}
    \includegraphics[width=\textwidth, trim=0 0 0 0, clip=true]{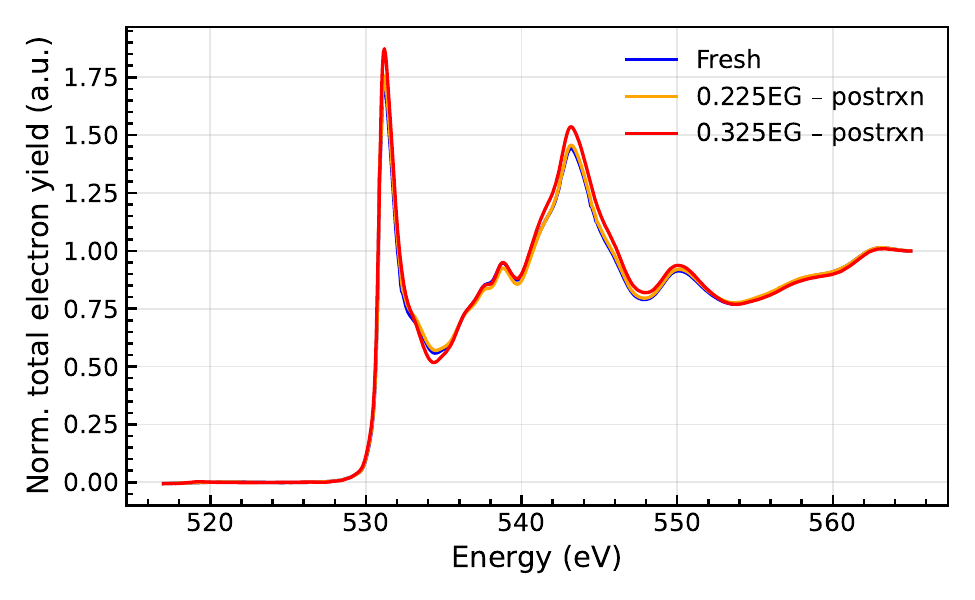}
\end{subfigure}%
\hfill
\begin{subfigure}{0.48\textwidth}
    \centering
    \caption{Theory}
    \label{fig:O_XAS_theory}
    \includegraphics[width=\textwidth, trim=0 0 0 0, clip=true]{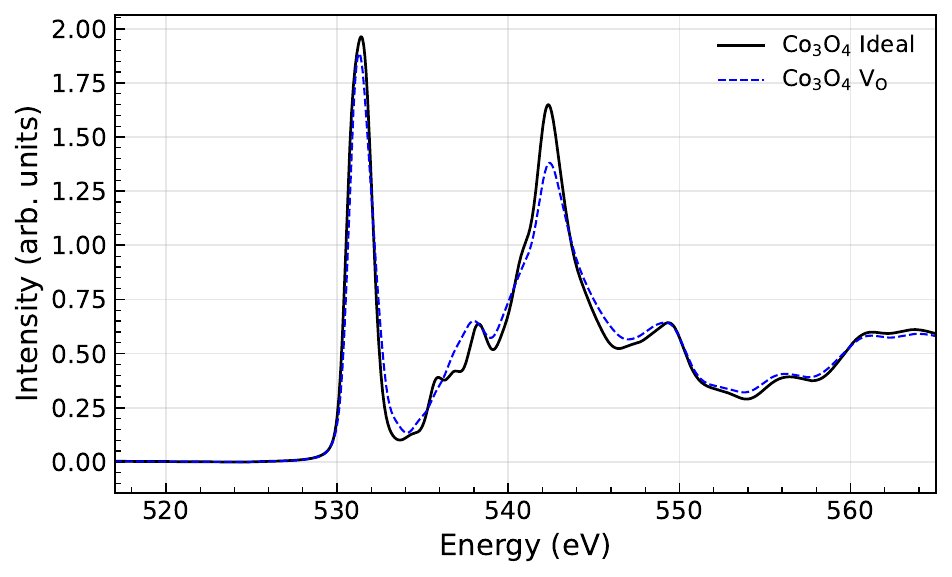}
\end{subfigure}
\caption{\justifying
O~K-edge X-ray absorption spectra (XAS) of Co$_3$O$_4$ for the hard-templated sample (HT) before and after liquid-phase ethylene glycol oxidation, compared with theoretical reference spectra. Panel~(a) shows the experimental spectra for the fresh catalyst (blue), the $0.025$~EG post-reaction state (orange), and the $0.325$~EG post-reaction state (red). Panel~(b) displays the theoretical O~K-edge spectra for the ideal Co$_3$O$_4$ spinel (black) and for a structure containing a single oxygen vacancy (blue). The reduced intensity of the first two peaks in the vacancy-containing system closely resembles the experimental fresh-state spectrum, while the ideal calculated spectrum matches the oxidized $0.325$~EG post-reaction data. The same spectral trend is observed for the SF and CP samples (see Figure~S1), all of which indicates that oxygen-vacancy-type defects are healed.}
\label{fig:O_XAS}
\end{figure*}

Figure~\ref{fig:O_XAS} shows the evolution of the O K-edge of Co$_3$O$_4$ for the hard-templated sample (HT) and how this evolution compares with our theoretical reference spectra. Panel (a) corresponds to the HT sample and contains three curves: the fresh catalyst (blue), the $0.025$ EG post reaction spectrum (orange), and the $0.325$ EG post reaction spectrum (red). Panel (b) shows the theoretical O K-edge for the ideal Co$_3$O$_4$ structure (black) and for Co$_3$O$_4$ containing a single oxygen vacancy (blue). Although the discussion in this work focuses on the HT sample, the corresponding experimental spectra for the spray flame synthesized (SF) and co--precipitated (CP) samples, which exhibit the same trends, are also provided in the Supporting Information (Figure S1).

For the HT sample the trend is the same as for the other two catalysts. The fresh material has a slightly weaker intensity at the onset of the absorption, that is in the first and second features of the O K-edge at 531\,eV and 543\,eV, respectively. After reaction, the spectra sharpen and the intensities at both the 531\,eV peak and 543\,eV rise. This effect is most pronounced for the $0.325$~EG post reaction state. The SF and CP samples follow the same behavior, as shown in Figure~S1, therefore all three synthesis routes lead to catalysts that move in the same spectral direction when they are exposed to the liquid-phase ethylene glycol oxidation conditions.

The comparison with the calculated spectra in panel~(b) makes this interpretation quantitative. In the calculation, the ideal Co$_3$O$_4$ has two relatively strong and well separated low energy features, because the O $2p$ states are strongly hybridized with Co $3d$ states in a fully oxidized spinel environment. When an oxygen vacancy is introduced, the theoretical spectrum loses intensity in exactly those first two features. 

The experimental fresh spectrum of HT in panel~(a), together with the fresh spectra of SF and CP in Figure~S1, resemble this vacancy-like calculated curve, while the $0.325$~EG post reaction spectra resemble much more the ideal calculated curve with stronger first and second peaks at 531\,eV and 543\,eV, respectively. In other words, the catalytically treated samples evolve from a vacancy-perturbed O K-edge toward an almost ideal spinel-like O K-edge. Therefore, one may deduce that this correspondence between experiment and theory supports that the liquid-phase reaction heals oxygen-vacancy-type defects and restores a more oxidized Co--O electronic structure. In other words, it can be suggested that there is a redox competition: ethylene glycol reduces the surface, while O$_2$ in the reaction atmosphere reoxidizes it. This is further supported by the observation that higher O$_2$ pressure leads to higher conversion, and that the catalyst remains relatively stable and reusable over multiple reaction cycles (see Section~\ref{sec:Catalysis}).

\textbf{Co K-edge X-ray Absorption Spectroscopy:}

\begin{figure*}[ht]
\centering
\includegraphics[width=0.6\textwidth, trim=0 0 0 0, clip=true]{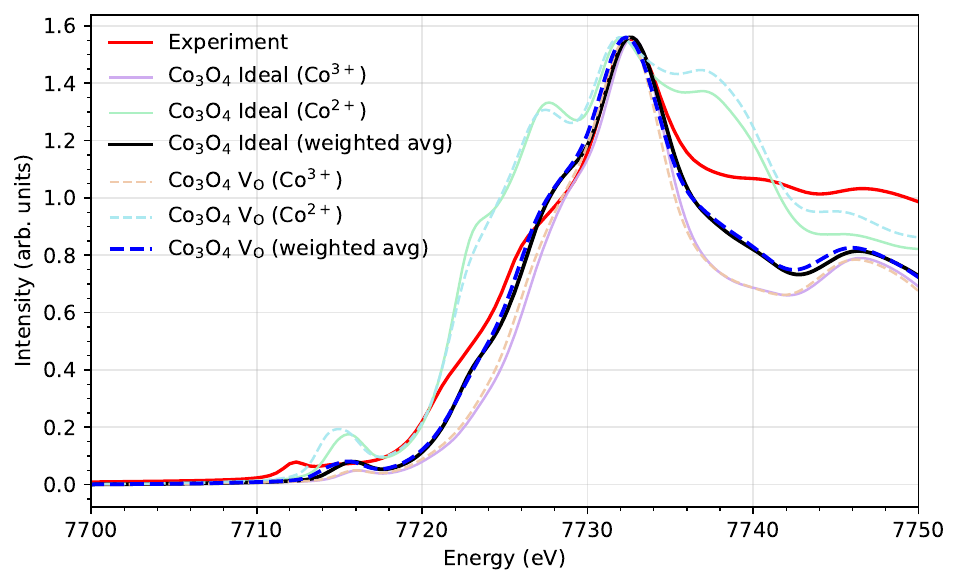}
\caption{\justifying
Co K-edge XANES of Co$_3$O$_4$ compared with site-resolved theoretical spectra. The experimental spectrum of the fresh hard-templated sample\cite{fandre2025unveiling} (red) serves as the reference. Calculated spectra for octahedral Co$^{3+}$ and tetrahedral Co$^{2+}$, along with their stoichiometric weighted average, reproduce the experimental spectra with excellent agreement. For the Co$_3$O$_4$ with the V$_\mathrm{O}$, the weighted spectrum exhibits a slight shift toward lower energy and a less steep edge, which suggests an increased Co$^{2+}$ character. This modest but consistent change supports the interpretation of vacancy-induced local reduction, in agreement with density of states (DOS), charge density difference (CDD), and O K-edge analyses as previously discussed.}
\label{fig:Co_XAS}
\end{figure*}

Figure~\ref{fig:Co_XAS} shows the Co K-edge XANES of Co$_3$O$_4$ together with the site-resolved calculated spectra and their weighted sums. The experimental spectrum of the fresh hard-templated sample\cite{fandre2025unveiling} is shown in red and serves as the reference (Co K-edge XAS for the post-reaction samples are not yet available at the time of this publication). The calculated spectra for octahedral Co$^{3+}$ and tetrahedral Co$^{2+}$ in the \emph{ideal} spinel, together with their stoichiometric weighted average, reproduce very well the edge position and the shape of the main rising feature in the experiment. As expected, the Co$^{2+}$ curve is slightly lower in energy, the Co$^{3+}$ curve is slightly higher and steeper, and the ``ideal'' (weighted avg) spectrum lies between them, approaching the experimental curve.

The same site-by-site construction was then carried out for the structure containing an oxygen vacancy. In this case, the ``V$_\mathrm{O}$'' (weighted avg) spectrum is slightly shifted toward lower energy and the edge is a bit less steep. Both effects are what one would expect if the vacancy locally increases the Co$^{2+}$ character. The change, however, is modest: the V$_\mathrm{O}$ and ideal weighted spectra remain close, but the direction of the change is consistent with the density of states (DOS), charge density difference (CDD), and O K-edge results, which all indicate a vacancy-induced local reduction. In other words, the way the V$_\mathrm{O}$ spectrum shifts to slightly lower energy while getting broader, is fully consistent with the previously suggested picture that an oxygen vacancy leads to more Co$^{2+}$ character.

\subsubsection{Magnetometry}\label{sec:Magnetometry}
In addition, magnetometry measurements have been performed on the hard-templated sample for the fresh and post-catalysis samples as a function of temperature (see Fig.~S3). Using the Curie-Weiss law, the temperature-dependent magnetization allows the calculation of the Curie constant and therefore the effective magnetic moment of the sample $\mu_\mathrm{eff}$ as described in Mugiraneza et al. \cite{Mugiraneza2022}. Between the fresh sample and the two post-catalysis samples an average reduction of 5\% of $\mu_\mathrm{eff}$ was observed. Considering that in $\mathrm{Co_3O_4}$ only Co$^{2+}$ contributes to the effective magnetic moment, the magnetometry results indicate that the Co$^{2+}$ amount is reduced after catalysis, which is in agreement with the statements in the previous section regarding oxygen refilling and the oxidation of Co$^{2+}$ to Co$^{3+}$.

\section{Conclusions}\label{sec:conclusions}

In the present work, we investigated oxygen vacancies in Co$_3$O$_4$ both in the bulk phase and under liquid-phase ethylene glycol oxidation by combining spin-polarized DFT+$U$, \textit{ab initio} molecular dynamics, and theoretical and experimental X-ray absorption spectroscopy (O and Co K-edge). Our main findings can be summarized as follows:

\begin{enumerate}
\item Introducing a single V$_\mathrm{O}$ into ideal bulk Co$_3$O$_4$ reduces two neighboring octahedral Co$^{3+}$ ions to high-spin Co$^{2+}$ and narrows the band gap.
\item These newly formed Co$^{2+}$ ions adopt similar local magnetic moments ($\sim 2.5~\mu_\mathrm{B}$) but in opposite directions.
\item They do not move into regular tetrahedral positions but remain in vacancy-perturbed, distorted octahedral environments with five-fold coordination.
\item \textit{Ab initio} MD at 300~K (50~ps) shows that both vacancy-induced Co$^{2+}$ sites remain stable, structurally (the Co--O coordination stays at 5 with only brief fluctuations) and magnetically (the local moments stay in the $\sim 2.5~\mu_\mathrm{B}$ Co$^{2+}$ range). This suggests that any further structural transformation would require overcoming a higher activation barrier than what is accessible at 300~K in our simulations.
\item Comparison of the computed O K-edge spectra for ideal Co$_3$O$_4$ and for Co$_3$O$_4$ with V$_\mathrm{O}$ to the experimental O K-edge spectra of the as-prepared and post-reaction catalysts shows that the fresh samples resemble the vacancy-containing calculation, while the spectra after liquid-phase ethylene glycol oxidation shift toward the ideal computed spectrum. This suggests that the reaction conditions tend to refill or heal oxygen-vacancy-type defects and drive the catalyst back toward a more oxidized, spinel-like state.
\item Consistently, magnetometry shows an average reduction of $5\%$ in effective magnetic moment after catalysis, suggesting a lower Co$^{2+}$ content.
\item This picture is further supported by the fact that higher O$_2$ pressures enhance EG conversion and that the catalyst remains stable and active across multiple reaction cycles.
\end{enumerate}

\section*{Acknowledgments}
We are grateful for funding by the Deutsche Forschungsgemeinschaft (DFG, German Research Foundation) in CRC/TRR 247 (project-ID 388390466: projects A01, A10 and B02) and under Germany's Excellence Strategy – EXC 2033 RESOLV (project-ID 390677874). The authors also acknowledge the computing time provided to them by the Paderborn Center for Parallel Computing (PC2). The authors thank the Helmholtz-Zentrum Berlin für Materialien und Energie for the allocation of synchrotron radiation beamtime within proposal ST-242-12862 at the synchrotron BESSY II and financial support for travel to BESSY II. Special thanks to Leon Müller and Christof Schulz for the spray-flame synthesized samples. Similarly to Carsten Placke-Yan and Stephan Schulz for the synthesis of co-precipitation samples. We acknowledge SOLEIL for the provision of synchrotron radiation facilities and would like to thank Dr. G. Landrot and the other beamline scientists for their assistance in using the SAMBA beamline (proposal no. 20240233).

\section*{Data Availability}
The data generated and analyzed during the current study are available from the corresponding authors upon request.
\bibliography{literature}

\bibliographystyle{achemso}


\end{document}


\setstretch{1.0}
\title{Supplementary Information: A Combined Theoretical and Experimental Study of Oxygen Vacancies in Co$_3$O$_4$ for Liquid-Phase Oxidation Catalysis}

\author{Amir Omranpour}
\email{\textcolor{black}{amir.omranpour@rub.de}}
\affiliation{Lehrstuhl f\"ur Theoretische Chemie II, Ruhr-Universit\"at Bochum, 44780 Bochum, Germany}
\affiliation{Research Center Chemical Sciences and Sustainability, Research Alliance Ruhr, 44780 Bochum, Germany}

\author{Lea K\"ammerer}
\affiliation{Faculty of Physics and Center for Nanointegration Duisburg-Essen (CENIDE), University of Duisburg–Essen, 47057 Duisburg, Germany}

\author{Catalina Leiva--Leroy}
\affiliation{Laboratory of Industrial Chemistry, Ruhr-Universit\"at Bochum, 44780 Bochum, Germany}

\author{Anna Rabe}
\affiliation{Faculty of Physics and Center for Nanointegration Duisburg-Essen (CENIDE), University of Duisburg–Essen, 47057 Duisburg, Germany}

\author{Takuma Sato}
\affiliation{Max Planck Institute for Chemical Energy Conversion, 45470 M\"ulheim an der Ruhr, Germany}

\author{Soma Salamon}
\affiliation{Faculty of Physics and Center for Nanointegration Duisburg-Essen (CENIDE), University of Duisburg–Essen, 47057 Duisburg, Germany}

\author{Joachim Landers}
\affiliation{Faculty of Physics and Center for Nanointegration Duisburg-Essen (CENIDE), University of Duisburg–Essen, 47057 Duisburg, Germany}

\author{Benedikt Eggert}
\affiliation{Faculty of Physics and Center for Nanointegration Duisburg-Essen (CENIDE), University of Duisburg–Essen, 47057 Duisburg, Germany}

\author{Eugen Weschke}
\affiliation{Helmholtz-Zentrum Berlin für Materialien und Energie (HZB), 12489 Berlin, Germany}

\author{Jean Pascal Fandr\'e}
\affiliation{Heterogeneous Catalysis and Sustainable Energy, Max-Planck-Institut für Kohlenforschung, Kaiser-Wilhelm-Platz 1, 45470 Mülheim an der Ruhr, Germany}

\author{Ashwani Kumar}
\affiliation{Heterogeneous Catalysis and Sustainable Energy, Max-Planck-Institut für Kohlenforschung, Kaiser-Wilhelm-Platz 1, 45470 Mülheim an der Ruhr, Germany}

\author{Harun T\"uys\"uz}
\affiliation{Heterogeneous Catalysis and Sustainable Energy, Max-Planck-Institut für Kohlenforschung, Kaiser-Wilhelm-Platz 1, 45470 Mülheim an der Ruhr, Germany}
\affiliation{Catalysis and Energy Materials Group, IMDEA Materials Institute, Calle Eric Kandel 2, 28906, Getafe, Madrid, Spain}

\author{Martin Muhler}
\affiliation{Laboratory of Industrial Chemistry, Ruhr-Universit\"at Bochum, 44780 Bochum, Germany}

\author{Heiko Wende}
\affiliation{Faculty of Physics and Center for Nanointegration Duisburg-Essen (CENIDE), University of Duisburg–Essen, 47057 Duisburg, Germany}

\author{J\"org Behler}
\affiliation{Lehrstuhl f\"ur Theoretische Chemie II, Ruhr-Universit\"at Bochum, 44780 Bochum, Germany}
\affiliation{Research Center Chemical Sciences and Sustainability, Research Alliance Ruhr, 44780 Bochum, Germany}

\date{\today}

\maketitle

\section{X-ray Absorption Spectroscopy}\label{sec:XAS}\label{sec:XAS}

\begin{figure*}[ht!]
\centering
\begin{subfigure}{0.48\textwidth}
    \centering
    \caption{Experiment (Sample SF)}
    \label{fig:O_XAS_SF}
    \includegraphics[width=\textwidth, trim=0 0 0 0, clip=true]{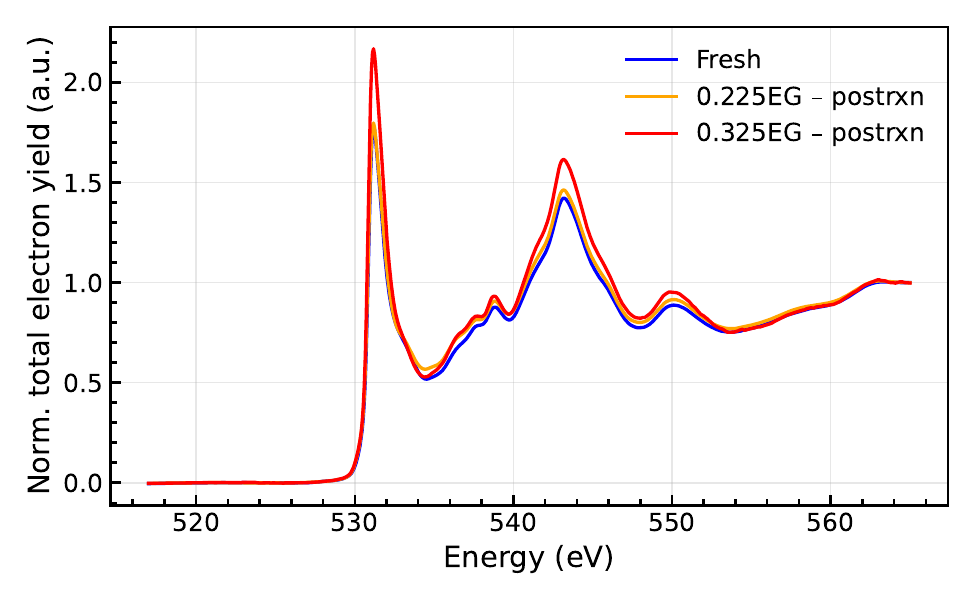}
\end{subfigure}%
\hfill
\begin{subfigure}{0.48\textwidth}
    \centering
    \caption{Experiment (Sample CP)}
    \label{fig:O_XAS_CP}
    \includegraphics[width=\textwidth, trim=0 0 0 0, clip=true]{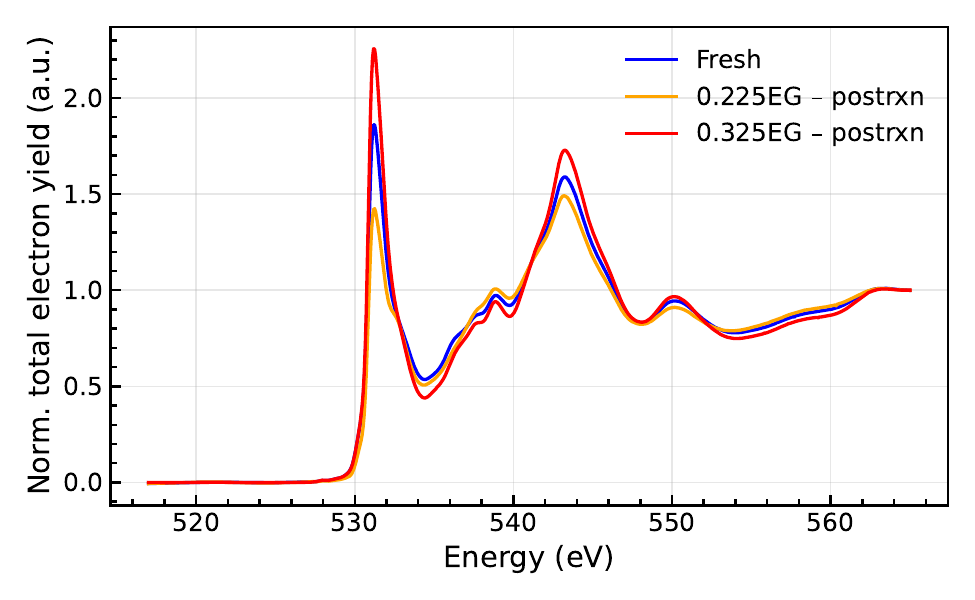}
\end{subfigure}
\caption{\justifying
O~K-edge X-ray absorption spectra (XAS) of Co$_3$O$_4$ for the spray-flame synthesized sample SF (panel~(a)) and the co--precipitated sample CP (panel~(b)) before and after liquid-phase ethylene glycol oxidation. Each panel includes spectra for the fresh catalyst (blue), the $0.025$~EG post-reaction state (orange), and the $0.325$~EG post-reaction state (red). In both cases, the $0.325$~EG post-reaction spectra exhibit stronger and sharper first and second peaks, indicating a more oxidized Co--O environment, consistent with the behaviour of the HT sample shown in Figure~6.}
\label{fig:O_XAS_SF_CP}
\end{figure*}

Figure~\ref{fig:O_XAS_SF_CP} presents additional O K-edge X-ray absorption spectra for the spray flame synthesized (SF) and co--precipitated (CP) Co$_3$O$_4$ samples before and after liquid phase ethylene glycol oxidation. The observed spectral changes closely follow those of the hard-templated sample discussed in the main text and confirm that the increased intensity and sharpening of the first two peaks at $0.325$~EG indicate a more oxidized Co--O environment across all synthesis routes.

\section{X-ray magnetic circular dichroism (XMCD)}\label{sec:XAS}\label{sec:XMCD}

\begin{figure*}[ht!]
\centering
\includegraphics[width=0.5\textwidth, trim=0 0 0 0, clip=true]{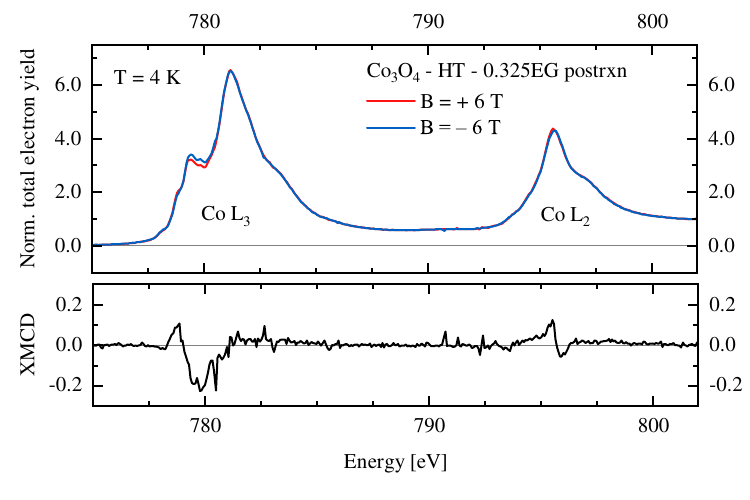}
\caption{\justifying
Co $L_{2,3}$-edges X-ray absorption near-edge spectra (top) and corresponding X-ray magnetic circular dichroism (bottom) of Co$_3$O$_4$ hard-templated (HT) sample after liquid-phase ethylene glycol oxidation with a concentration of 0.325 EG measured at 4\,K and an applied magnetic field of 6\,T. }
\label{fig:Co_XMCD}
\end{figure*}

Figure~\ref{fig:Co_XMCD} exemplarily shows the XANES measured at the Co $L_{2,3}$-edges with the corresponding XMCD for the Co$_3$O$_4$ hard-templated (HT) sample after liquid-phase ethylene glycol oxidation with a concentration of 0.325 EG measured at 4\,K and an applied magnetic field of 6\,T. As suggested by the DFT calculations for the O K-edge, the Co$^{3+}$ ions are reduced to Co$^{2+}$ ions located at the octahedral sites due to oxygen vacancies. Therefore, it can be suggested that uncompensated moments arise as the Co$^{2+}$ ions adopt a high-spin configuration which leads to a finite XMCD signal, as exemplified in Fig.~\ref{fig:Co_XMCD}.

\section{Magnetometry measurements}\label{magnetometry}

\begin{figure*}[ht!]
\centering
\begin{subfigure}{0.48\textwidth}
    \centering
    \caption{Temperature-dependent mass magnetization}
    \includegraphics[width=\textwidth, trim=0 0 0 0, clip=true]{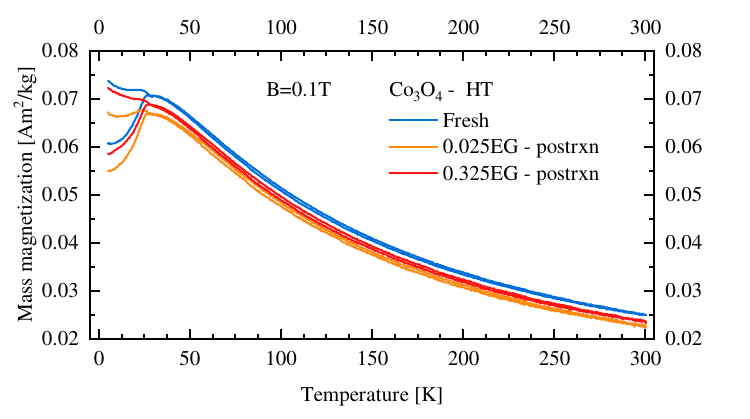}
    \label{fig:magnetometry_a}
\end{subfigure}%
\hfill
\begin{subfigure}{0.48\textwidth}
    \centering
    \caption{Curie-Weiss law fit}
    \includegraphics[width=\textwidth, trim=0 0 0 0, clip=true]{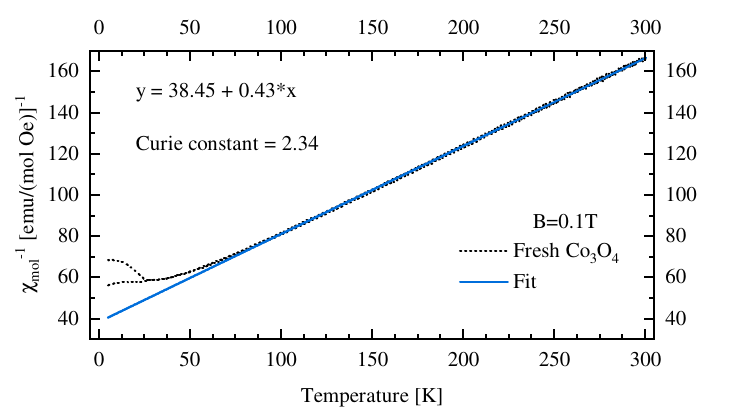}
    \label{fig:magnetometry_b}
\end{subfigure}
\caption{\justifying
(a) Temperature-dependent mass magnetization of the three hard-templated samples: fresh (blue), 0.025 EG post-reaction (orange), and 0.325 EG post-reaction (red) measured with an applied magnetic field of 0.1\,T in a temperature range from 5 to 300\,K. (b) Exemplary inverse molar susceptibility $\chi_\mathrm{mol}$ as a function of temperature (black dotted line) with a linear fit to obtain the Curie constant (blue) for the fresh sample.}
\label{fig:magnetometry}
\end{figure*}

Figure~\ref{fig:magnetometry} (a) shows the temperature-dependent mass magnetization in a range from 5-300\,K for the three hard-templated samples before (blue) and after liquid phase ethylene glycol oxidation (0.025 EG post-reaction (orange) and 0.325 EG post-reaction (red)) under an applied magnetic field of 0.1\,T. Using the inverse molar susceptibility $\chi_\mathrm{mol}$ calculated via equation 7 in \cite{Mugiraneza2022} it is possible to obtain the effective magnetic moment $\mu_\mathrm{eff}$ via the Curie constant by fitting with a linear equation presented in Fig.~\ref{fig:magnetometry} (b) for the fresh sample. The fit parameters given in panel (b) are rounded to the second decimal. The obtained $\mu_\mathrm{eff}$ show an average reduction of 5\% when comparing the fresh sample with the post-catalysis ones ($\mu_\mathrm{eff} = 4.33 \mu_\mathrm{B}$ (Fresh) vs. $\mu_\mathrm{eff} = 4.06 \mu_\mathrm{B}$ (0.025 EG post-reaction) and $\mu_\mathrm{eff} = 4.17 \mu_\mathrm{B}$ (0.325 EG post-reaction)) following the observation of the theoretical calculations that the Co$^{2+}$ amount is reduced after catalysis under the assumption that only Co$^{2+}$ contributes to $\mu_\mathrm{eff}$ in $\mathrm{Co_3O_4}$.

\section{Site-resolved magnetic moments from DFT}\label{sec:DFT}

Tables~S1–S3 show the details of the orbital-resolved spin moments (s, p, d) and the resulting total local magnetic moment for every ion in the Co$_3$O$_4$ supercell obtained from our DFT calculations.

In all tables, the atoms follow the same ordering:
\begin{itemize}
    \item ions 1–8 correspond to the tetrahedral Co$^{2+}$ sites,
    \item ions 9–24 correspond to the octahedral Co$^{3+}$ sites,
    \item the remaining ions are the O atoms.
\end{itemize}
This ordering makes it straightforward to assign the expected $\sim 2.5$–$2.6~\mu_\mathrm{B}$ moments on Co$^{2+}$, the nearly quenched moments on Co$^{3+}$, and the small induced moments on oxygen.

Table~S1 shows the reference (ideal) Co$_3$O$_4$ calculation without an oxygen vacancy. As expected for the normal spinel, the first eight Co atoms (tetrahedral Co$^{2+}$) carry moments of about $\pm 2.58~\mu_\mathrm{B}$, while the following sixteen Co atoms (octahedral Co$^{3+}$) have moments very close to zero. This confirms the low-spin Co$^{3+}$ configuration. Oxygen atoms exhibit only small polarization (on the order of $10^{-2}~\mu_\mathrm{B}$), and the total moment of the cell is essentially zero which is consistent with the antiferromagnetic arrangement.

Table~S2 reports the same site-resolved moments for the structure containing a single oxygen vacancy. In this case, two of the octahedral Co$^{3+}$ ions adjacent to the vacancy develop finite moments in the Co$^{2+}$ range, indicating the local reduction of Co$^{3+}$ to Co$^{2+}$ induced by V$_\mathrm{O}$. The rest of the Co sites keep the values they had in the ideal structure, and only small additional polarization appears on nearby O atoms. The total magnetic moment of the supercell remains close to zero, because the two newly formed Co$^{2+}$ moments couple antiferromagnetically.

Table~S3 shows the ideal Co$_3$O$_4$ case recalculated  by taking into account the spin–orbit coupling (SOC) contribution. The overall pattern is the same as in Table~S1: the first eight Co$^{2+}$ sites retain large moments, the octahedral Co$^{3+}$ sites stay nearly nonmagnetic, and the oxygens show only minor induced polarization. SOC introduces only small quantitative changes to the individual site moments and does not modify the magnetic ordering.

\begin{table*}[ht]
\centering
\caption{\justifying
Site-resolved orbital contributions (s, p, d) and total local magnetic moments for the ideal Co$_3$O$_4$ supercell from spin-polarized DFT(+U). Ions 1–8 are tetrahedral Co$^{2+}$, ions 9–24 are octahedral Co$^{3+}$, and the remaining ions are O. Co$^{2+}$ sites carry moments of $\sim 2.6~\mu_\mathrm{B}$, while Co$^{3+}$ sites are nearly nonmagnetic.}
\begin{tabular}{rcccc}
\toprule
\# of ion & $s$ & $p$ & $d$ & total \\
\midrule
1  &  0.013 &  0.050 &  2.518 &  2.581 \\
2  &  0.013 &  0.050 &  2.518 &  2.581 \\
3  &  0.013 &  0.050 &  2.518 &  2.581 \\
4  &  0.013 &  0.050 &  2.518 &  2.581 \\
5  & -0.013 & -0.050 & -2.518 & -2.581 \\
6  & -0.013 & -0.050 & -2.518 & -2.581 \\
7  & -0.013 & -0.050 & -2.518 & -2.581 \\
8  & -0.013 & -0.050 & -2.518 & -2.581 \\
9  & -0.000 & -0.000 &  0.000 & -0.000 \\
10 & -0.000 &  0.000 & -0.000 & -0.000 \\
11 &  0.000 & -0.000 &  0.000 &  0.000 \\
12 &  0.000 &  0.000 & -0.000 &  0.000 \\
13 &  0.000 &  0.000 &  0.000 &  0.000 \\
14 & -0.000 &  0.000 &  0.000 &  0.000 \\
15 & -0.000 &  0.000 & -0.000 & -0.000 \\
16 & -0.000 &  0.000 & -0.000 & -0.000 \\
17 &  0.000 & -0.000 &  0.000 &  0.000 \\
18 &  0.000 &  0.000 &  0.000 &  0.000 \\
19 & -0.000 &  0.000 &  0.000 &  0.000 \\
20 & -0.000 & -0.000 & -0.000 & -0.000 \\
21 &  0.000 &  0.000 & -0.000 & -0.000 \\
22 &  0.000 &  0.000 & -0.000 &  0.000 \\
23 &  0.000 &  0.000 &  0.000 &  0.000 \\
24 &  0.000 &  0.000 &  0.000 &  0.000 \\
25 & -0.004 & -0.029 &  0.000 & -0.033 \\
26 &  0.004 &  0.029 &  0.000 &  0.033 \\
27 & -0.004 & -0.029 &  0.000 & -0.033 \\
28 &  0.004 &  0.029 &  0.000 &  0.033 \\
29 &  0.004 &  0.029 &  0.000 &  0.033 \\
30 & -0.004 & -0.029 &  0.000 & -0.033 \\
31 &  0.004 &  0.029 &  0.000 &  0.033 \\
32 & -0.004 & -0.029 &  0.000 & -0.033 \\
33 & -0.004 & -0.029 &  0.000 & -0.033 \\
34 &  0.004 &  0.029 &  0.000 &  0.033 \\
35 &  0.004 &  0.029 &  0.000 &  0.033 \\
36 & -0.004 & -0.029 &  0.000 & -0.033 \\
37 & -0.004 & -0.029 &  0.000 & -0.033 \\
38 &  0.004 &  0.029 &  0.000 &  0.033 \\
39 &  0.004 &  0.029 &  0.000 &  0.033 \\
40 & -0.004 & -0.029 &  0.000 & -0.033 \\
41 &  0.004 &  0.029 &  0.000 &  0.033 \\
42 & -0.004 & -0.029 &  0.000 & -0.033 \\
43 &  0.004 &  0.029 &  0.000 &  0.033 \\
44 & -0.004 & -0.029 &  0.000 & -0.033 \\
45 & -0.004 & -0.029 &  0.000 & -0.033 \\
46 &  0.004 &  0.029 &  0.000 &  0.033 \\
47 & -0.004 & -0.029 &  0.000 & -0.033 \\
48 &  0.004 &  0.029 &  0.000 &  0.033 \\
49 &  0.004 &  0.029 &  0.000 &  0.033 \\
50 & -0.004 & -0.029 &  0.000 & -0.033 \\
51 & -0.004 & -0.029 &  0.000 & -0.033 \\
52 &  0.004 &  0.029 &  0.000 &  0.033 \\
53 &  0.004 &  0.029 &  0.000 &  0.033 \\
54 & -0.004 & -0.029 &  0.000 & -0.033 \\
55 & -0.004 & -0.029 &  0.000 & -0.033 \\
56 &  0.004 &  0.029 &  0.000 &  0.033 \\
\textbf{Total} &  0.000 & -0.000 & 0.000 & -0.000 \\
\bottomrule
\end{tabular}
\end{table*}

\begin{table*}[ht]
\centering
\caption{\justifying
Site-resolved orbital contributions (s, p, d) and total local magnetic moments for the Co$_3$O$_4$ supercell containing a single oxygen vacancy V$_\mathrm{O}$. Ions 1–8 are tetrahedral Co$^{2+}$, ions 9–24 are octahedral Co$^{3+}$ in the ideal structure, and the remaining ions are O. Two of the octahedral Co$^{3+}$ sites adjacent to the vacancy acquire finite moments in the Co$^{2+}$ range, indicating local reduction induced by V$_\mathrm{O}$.}
\begin{tabular}{rcccc}
\toprule
\# of ion & $s$ & $p$ & $d$ & total \\
\midrule
1  &  0.014 &  0.048 &  2.515 &  2.577 \\
2  &  0.013 &  0.050 &  2.510 &  2.573 \\
3  &  0.013 &  0.048 &  2.521 &  2.582 \\
4  &  0.013 &  0.047 &  2.521 &  2.581 \\
5  & -0.013 & -0.048 & -2.524 & -2.585 \\
6  & -0.014 & -0.049 & -2.524 & -2.587 \\
7  & -0.013 & -0.048 & -2.522 & -2.583 \\
8  & -0.031 & -0.044 & -2.488 & -2.563 \\
9  &  0.016 &  0.023 &  2.412 &  2.451 \\
10 & -0.014 & -0.021 & -2.448 & -2.484 \\
11 &  0.001 &  0.002 &  0.035 &  0.038 \\
12 &  0.000 &  0.000 &  0.011 &  0.011 \\
13 & -0.000 & -0.000 & -0.011 & -0.012 \\
14 &  0.000 &  0.000 &  0.002 &  0.002 \\
15 &  0.000 & -0.000 & -0.009 & -0.009 \\
16 &  0.001 &  0.001 &  0.010 &  0.012 \\
17 & -0.000 &  0.000 &  0.011 &  0.012 \\
18 & -0.000 & -0.001 &  0.002 &  0.001 \\
19 &  0.000 & -0.000 &  0.003 &  0.003 \\
20 & -0.000 & -0.000 & -0.001 & -0.001 \\
21 &  0.000 &  0.000 &  0.021 &  0.022 \\
22 &  0.000 & -0.000 & -0.004 & -0.003 \\
23 & -0.000 & -0.001 & -0.005 & -0.005 \\
24 & -0.000 &  0.000 & -0.001 & -0.001 \\
25 & -0.004 & -0.029 &  0.000 & -0.033 \\
26 &  0.005 &  0.029 &  0.000 &  0.033 \\
27 & -0.004 & -0.029 &  0.000 & -0.033 \\
28 &  0.005 &  0.023 &  0.000 &  0.028 \\
29 &  0.004 &  0.026 &  0.000 &  0.030 \\
30 & -0.004 & -0.028 &  0.000 & -0.032 \\
31 &  0.004 &  0.031 &  0.000 &  0.035 \\
32 & -0.004 & -0.028 &  0.000 & -0.032 \\
33 & -0.004 & -0.041 &  0.000 & -0.045 \\
34 &  0.000 & -0.002 &  0.000 & -0.002 \\
35 &  0.004 &  0.027 &  0.000 &  0.031 \\
36 & -0.004 & -0.026 &  0.000 & -0.030 \\
37 & -0.004 & -0.027 &  0.000 & -0.031 \\
38 &  0.008 &  0.074 &  0.000 &  0.082 \\
39 &  0.004 &  0.024 &  0.000 &  0.028 \\
40 & -0.003 & -0.041 &  0.000 & -0.044 \\
41 &  0.008 &  0.072 &  0.000 &  0.080 \\
42 & -0.005 & -0.043 &  0.000 & -0.048 \\
43 &  0.001 & -0.011 &  0.000 & -0.010 \\
44 & -0.003 & -0.039 &  0.000 & -0.042 \\
45 &  0.000 &  0.001 &  0.000 &  0.001 \\
46 & -0.004 & -0.029 &  0.000 & -0.033 \\
47 &  0.008 &  0.073 &  0.000 &  0.081 \\
48 &  0.004 &  0.021 &  0.000 &  0.025 \\
49 & -0.009 & -0.056 &  0.000 & -0.065 \\
50 & -0.004 & -0.027 &  0.000 & -0.031 \\
51 &  0.004 &  0.025 &  0.000 &  0.029 \\
52 &  0.004 &  0.025 &  0.000 &  0.029 \\
53 &  0.000 & -0.017 &  0.000 & -0.017 \\
54 & -0.004 & -0.028 &  0.000 & -0.032 \\
55 &  0.004 &  0.023 &  0.000 &  0.027 \\
\textbf{Total} & -0.007 & -0.021 & 0.038 & 0.010 \\
\bottomrule
\end{tabular}
\end{table*}

\begin{table*}[ht]
\centering
\caption{\justifying
Site-resolved orbital contributions (s, p, d) and total local magnetic moments for the ideal Co$_3$O$_4$ supercell recalculated with spin–orbit coupling (SOC) in the z direction. Ions 1–8 are tetrahedral Co$^{2+}$, ions 9–24 are octahedral Co$^{3+}$, and the remaining ions are O. SOC causes only minor quantitative changes and preserves the overall magnetic pattern of the ideal spinel.}
\begin{tabular}{rcccc}
\toprule
\# of ion & $s$ & $p$ & $d$ & total \\
\midrule
1  &  0.013 &  0.050 &  2.515 &  2.579 \\
2  &  0.013 &  0.050 &  2.515 &  2.579 \\
3  &  0.013 &  0.050 &  2.515 &  2.579 \\
4  &  0.013 &  0.050 &  2.515 &  2.579 \\
5  & -0.013 & -0.050 & -2.515 & -2.579 \\
6  & -0.013 & -0.050 & -2.515 & -2.579 \\
7  & -0.013 & -0.050 & -2.515 & -2.579 \\
8  & -0.013 & -0.050 & -2.515 & -2.579 \\
9  & -0.000 &  0.000 &  0.000 &  0.000 \\
10 & -0.000 &  0.000 &  0.000 &  0.000 \\
11 & -0.000 & -0.000 & -0.000 & -0.000 \\
12 &  0.000 &  0.000 &  0.000 &  0.000 \\
13 & -0.000 &  0.000 &  0.000 &  0.000 \\
14 &  0.000 &  0.000 &  0.000 &  0.000 \\
15 &  0.000 &  0.000 &  0.000 &  0.000 \\
16 &  0.000 & -0.000 & -0.000 & -0.000 \\
17 & -0.000 & -0.000 & -0.000 & -0.000 \\
18 & -0.000 &  0.000 &  0.000 &  0.000 \\
19 & -0.000 &  0.000 &  0.000 &  0.000 \\
20 &  0.000 & -0.000 & -0.000 & -0.000 \\
21 & -0.000 &  0.000 &  0.000 &  0.000 \\
22 &  0.000 &  0.000 &  0.000 &  0.000 \\
23 &  0.000 &  0.000 &  0.000 &  0.000 \\
24 & -0.000 &  0.000 &  0.000 &  0.000 \\
25 & -0.004 & -0.029 &  0.000 & -0.033 \\
26 &  0.004 &  0.029 &  0.000 &  0.033 \\
27 & -0.004 & -0.029 &  0.000 & -0.033 \\
28 &  0.004 &  0.029 &  0.000 &  0.033 \\
29 &  0.004 &  0.029 &  0.000 &  0.033 \\
30 & -0.004 & -0.029 &  0.000 & -0.033 \\
31 &  0.004 &  0.029 &  0.000 &  0.033 \\
32 & -0.004 & -0.029 &  0.000 & -0.033 \\
33 & -0.004 & -0.029 &  0.000 & -0.033 \\
34 &  0.004 &  0.029 &  0.000 &  0.033 \\
35 &  0.004 &  0.029 &  0.000 &  0.033 \\
36 & -0.004 & -0.029 &  0.000 & -0.033 \\
37 & -0.004 & -0.029 &  0.000 & -0.033 \\
38 &  0.004 &  0.029 &  0.000 &  0.033 \\
39 &  0.004 &  0.029 &  0.000 &  0.033 \\
40 & -0.004 & -0.029 &  0.000 & -0.033 \\
41 &  0.004 &  0.029 &  0.000 &  0.033 \\
42 & -0.004 & -0.029 &  0.000 & -0.033 \\
43 &  0.004 &  0.029 &  0.000 &  0.033 \\
44 & -0.004 & -0.029 &  0.000 & -0.033 \\
45 & -0.004 & -0.029 &  0.000 & -0.033 \\
46 &  0.004 &  0.029 &  0.000 &  0.033 \\
47 & -0.004 & -0.029 &  0.000 & -0.033 \\
48 &  0.004 &  0.029 &  0.000 &  0.033 \\
49 &  0.004 &  0.029 &  0.000 &  0.033 \\
50 & -0.004 & -0.029 &  0.000 & -0.033 \\
51 & -0.004 & -0.029 &  0.000 & -0.033 \\
52 &  0.004 &  0.029 &  0.000 &  0.033 \\
53 &  0.004 &  0.029 &  0.000 &  0.033 \\
54 & -0.004 & -0.029 &  0.000 & -0.033 \\
55 & -0.004 & -0.029 &  0.000 & -0.033 \\
56 &  0.004 &  0.029 &  0.000 &  0.033 \\
\textbf{Total} & -0.000 & -0.001 & 0.001 & 0.001 \\
\bottomrule
\end{tabular}
\end{table*}

\bibliography{literature}

\bibliographystyle{achemso}